\documentclass[journal]{IEEEtran}
\usepackage{cite}
\usepackage{amsmath,amssymb,amsfonts}
\usepackage{algorithmic}
\usepackage{graphicx,color}
\usepackage{textcomp}
\usepackage{xcolor}
\usepackage{hyperref}
\hypersetup{hidelinks}
\usepackage{algorithm,algorithmic}
\usepackage{makecell}
\usepackage{multirow}
\usepackage{url}
\usepackage{array}
\usepackage{siunitx}
\usepackage{booktabs}
\usepackage{tabularx}
\usepackage[normalem]{ulem}
\usepackage{makecell}
\usepackage{soul}
\usepackage{float}
\usepackage{booktabs}
\usepackage{threeparttable}

\def\BibTeX{{\rm B\kern-.05em{\sc i\kern-.025em b}\kern-.08em
    T\kern-.1667em\lower.7ex\hbox{E}\kern-.125emX}}
\AtBeginDocument{\definecolor{tmlcncolor}{cmyk}{0.93,0.59,0.15,0.02}\definecolor{NavyBlue}{RGB}{0,86,125}}

\makeatletter
\let\arxivOriginalTitleStyle\ps@IEEEtitlepagestyle
\def\ps@IEEEtitlepagestyle{%
  \arxivOriginalTitleStyle
  \def\@oddhead{\parbox[b]{\textwidth}{\centering\normalfont\footnotesize
    Accepted for publication in \textit{IEEE Open Journal of Circuits and Systems}\\
    Accepted 25 July 2026\\
    \href{https://doi.org/10.1109/OJCAS.2026.3720349}{https://doi.org/10.1109/OJCAS.2026.3720349}\par}}%
  \let\@evenhead\@oddhead
}
\makeatother

\begin{document}

\markboth{}{Zhou {et al.}}

\title{FPGA Acceleration of CNN-BiLSTM for EEG-Based Sleep Staging Featuring FFT-Based Non-Unit Stride Fast Convolution}

\author{Aaron Zhou, 
        Yaqian~Xu, Graduate Student Member, IEEE,
        Chen~Chen, 
        James~Yen, 
        Weihang~Zheng, 
        Mark~Zhao, 
        Mingzhe~Sun, 
        Naize~Yang, 
        Andrew~G.~Richardson, Senior Member, IEEE,
        and Xilin~Liu, Senior Member, IEEE%
\thanks{Aaron Zhou, Yaqian Xu, Chen Chen, James Yen, Weihang Zheng, Mark Zhao, Mingzhe Sun, Naize Yang, and Xilin Liu are with the Edward S. Rogers Sr. Department of Electrical and Computer Engineering, University of Toronto, Toronto, ON M5G 2A2, Canada.}
\thanks{Andrew G. Richardson is with the Department of Neurosurgery, University of Pennsylvania, Philadelphia, PA 19104, USA.}
\thanks{Copyright 2026 The Authors. Licensed under \href{https://creativecommons.org/licenses/by/4.0/}{CC BY 4.0}.}}

\maketitle

\begin{abstract}
Deep learning (DL) models for real-time wearable sleep staging often rely on long-kernel temporal convolutions, a requirement that current accelerators struggle to support efficiently. To overcome this, we present an FPGA-based CNN-BiLSTM accelerator that incorporates several algorithm-hardware co-design techniques to enable efficient edge inference. Central to this design is an FFT-based non-unit stride fast convolution, which performs accumulation in the frequency domain to enable a single IFFT per block, reducing post-processing complexity to $O(N\log N)$. On the target long-kernel convolution layer, this yields a 20.5$\times$ reduction in arithmetic operations. Power-of-two quantization is applied across the pipeline, including FFT-based convolution and BiLSTM activations, {with an average accuracy degradation of only 0.11\% relative to} the floating-point baseline {across all subjects on the public MASS-SS3 dataset}. Additional optimizations, including block convolution, depth-first execution, and hardware engine reuse minimize memory access and enhance resource utilization. The accelerator, implemented on a Xilinx Zynq-7000 FPGA, operates at 10~MHz and consumes 11~mW of dynamic power including data handling and configuration overhead from a microcontroller. The design achieves a DSP power efficiency of 0.91~GOPS/W/DSP using only 4 DSPs, demonstrating scalable and energy-efficient DL acceleration for sleep staging that can be extended to other wearable applications.
\end{abstract}

\begin{IEEEkeywords}
Edge DL, EEG, sleep staging, CNN, BiLSTM, FFT-based convolution, non-unit stride, power-of-two quantization, FPGA
\end{IEEEkeywords}

\section{Introduction}

Sleep is essential for cognitive and physiological health, and sleep disorders are widespread and associated with severe health consequences~\cite{institute2006sleep}. Sleep modulation has emerged as a promising neuromodulation strategy for improving sleep quality and treating sleep-related disorders, but its effectiveness depends on accurate, real-time tracking of transitions among sleep stages, including NREM stages N1--N3, REM sleep, and wakefulness~\cite{o2014montreal}. Recent closed-loop paradigms employ real-time sleep stage classification (i.e., sleep staging) to deliver stage-specific stimulation, enabling causal investigation of sleep functions and therapeutic interventions, such as memory enhancement through stimulation during deep sleep~\cite{sun2023design,ngo2013auditory,xu2026inear}. A key requirement of such systems is automatic, low-latency sleep staging based on physiological signals such as EEG, which cannot rely on manual scoring used in offline studies~\cite{carroll2010polysomnography}.

Deep learning (DL) {has become the dominant approach for automatic sleep staging due to its ability to learn discriminative temporal features directly from physiological signals~\cite{mousavi2019sleepeegnet,kavoosi2023morpheusnet,supratak2017deepsleepnet}.} However, existing approaches are computationally intensive and often depend on multiple physiological signal channels, limiting their suitability for wearable, energy-constrained closed-loop systems~\cite{supratak2020tinysleepnet,SeqSleepNet,sun2023design}. Although lightweight models have been explored,
{their deployment on low-power wearable devices remains challenging because long-kernel convolutions dominate the computational workload, while recurrent inference introduces sequential data dependencies, nonlinear activations, and additional memory traffic~\cite{supratak2020tinysleepnet,wang2022review}.}

{
Existing hardware DL accelerators have primarily been developed for computer-vision and natural language processing workloads using small convolution kernels, where techniques such as Winograd or conventional FFT-based convolution are effective~\cite{kala2019high, liang2019evaluating}. However, these approaches do not efficiently support the large-kernel, non-unit-stride convolutions commonly found in EEG-based sleep-staging models. Furthermore, hardware support for recurrent networks often requires dedicated computation units that increase resource utilization and power consumption. Efficient acceleration of these networks therefore remains an important challenge for sleep staging, as well as many other EEG-based wearable applications.
}

To address these challenges, we present a DL accelerator for single-channel EEG-based sleep staging, with a focus on algorithm-hardware co-design for energy-efficient inference {and have fully validated our design on a Xilinx Zynq-7000 FPGA}. In contrast to prior work that emphasizes compact classifiers or generic CNN acceleration, this work targets a class of workloads commonly found in EEG-based applications that require long temporal context. Through FFT-based non-unit stride convolution and fixed-point recurrent inference with unified hardware engine reuse, this design enables practical closed-loop sleep staging under strict power and memory constraints. The key contributions of this work are summarized below.

\begin{itemize}
    \item \textbf{Efficient FFT-based acceleration of large-kernel convolutions with non-unit stride:} We propose an FFT-based fast convolution formulation for CNN layers with large kernel sizes (e.g., 1280) and non-unit strides (e.g., 12), which is not efficiently supported by conventional CNN accelerators or existing FFT-based acceleration techniques. By enabling frequency-domain accumulation and avoiding $O(SN\log N)$ post-processing computation, the proposed method reduces the arithmetic complexity of the dominant convolution layer by 20.5$\times$ while remaining hardware-efficient.
    \item \textbf{BiLSTM acceleration with unified hardware support and resource reuse:} {To support temporal context modeling essential for sleep staging, the accelerator reuses the same computation engine for CNN and BiLSTM operations, reducing hardware redundancy} and enabling concurrent operations. The BiLSTM engine also incorporates linear interpolation with minimal look-up operations for activation functions such as \texttt{tanh}, and shares look-up tables (LUTs) between \texttt{tanh} and \texttt{sigmoid} functions to further reduce hardware overhead.
    \item \textbf{{Algorithm-hardware} co-design for {memory- and power-efficient inference}:} We pre-process the model with block convolution and tiling to minimize memory footprint prior to inference. A depth-first execution scheme at the hardware level complements our strategies to {minimize external memory traffic, which is a primary contributor to energy consumption. These techniques are co-designed with the stride-based FFT formulation to enable efficient execution of long-kernel convolutions under strict power and memory constraints.}
    \item \textbf{Power-of-two (PoT) quantization supporting mixed FFT convolution and BiLSTM nonlinear activations:} We develop a PoT quantization scheme with scale selection configured for mixed conventional/FFT convolution and BiLSTM nonlinear activations. This enables efficient fixed-point arithmetic while suppressing accumulated numerical noise in large-point FFT-based convolution pipelines. The approach preserves numerical fidelity and simplifies hardware scaling operations, resulting in a negligible {0.11\% average} accuracy degradation relative to the floating-point baseline and providing a practical trade-off between numerical robustness and hardware simplicity.
\end{itemize}

The paper is organized as follows. Section~II introduces the fast convolution algorithm and the quantization scheme for the CNN-BiLSTM model. Section~III details the design and implementation of the FPGA accelerator with memory and power optimization techniques. Section~IV presents the experimental results. Section~V provides a comparative analysis and Section~VI concludes the paper.

\section{Hardware-Oriented Model Optimization Algorithms}

\subsection{CNN-BiLSTM Model}\label{sec:model}

{
Recent DL approaches for sleep staging can be broadly categorized into CNN–RNN, hierarchical sequence, attention-based, and fully convolutional architectures. CNN–RNN models such as DeepSleepNet~\cite{supratak2017deepsleepnet}, TinySleepNet~\cite{supratak2020tinysleepnet}, and IITNet~\cite{seo2020iitnet} combine temporal convolution with recurrent sequence modeling to jointly capture sleep morphology within an epoch and stage transitions across consecutive epochs, and have demonstrated competitive performance using single-channel EEG. Hierarchical sequence models, including SeqSleepNet~\cite{SeqSleepNet} and SleepEEGNet~\cite{mousavi2019sleepeegnet}, further model intra- and inter-epoch temporal dependencies through multi-level recurrent architectures. More recently, attention-based models such as AttnSleep~\cite{eldele2021attnsleep} and SleepTransformer~\cite{phan2022sleeptransformer} have replaced recurrent modules with self-attention to model long-range temporal context, while fully convolutional networks such as U-Sleep and resource-oriented architectures such as MorpheusNet have explored end-to-end convolutional inference or lightweight network redesign for embedded deployment~\cite{perslev2021usleep,kavoosi2023morpheusnet}. Despite these architectural differences, many state-of-the-art sleep-staging models share two important computational characteristics: long temporal convolutions for extracting sleep-specific waveforms and temporal modeling across consecutive epochs through recurrent or sequence-processing modules. These operations dominate computation, memory traffic, and hardware complexity, making them the principal bottlenecks for low-power edge deployment~\cite{wang2022review}.
}

{
The CNN-BiLSTM topology adopted in this work was previously proposed by our group in~\cite{sun2023design} and is representative of a broad class of modern sleep-staging models.
It combines convolutional feature extraction with BiLSTM-based temporal modeling while operating on a single F4-LER EEG channel. Previous studies have shown that single-channel EEG can effectively capture characteristic sleep events such as K-complexes and sleep spindles, achieving performance comparable to multi-channel approaches when combined with modern DL architectures~\cite{supratak2017deepsleepnet,supratak2020tinysleepnet,Single-channel-survey,zhang2024comparative}. This observation is further supported by the work of Korkalainen \textit{et al.}, who directly compared single-channel EEG with a combined EEG+EOG configuration and reported that incorporating EOG improved sleep-staging accuracy by only 0.2\%~\cite{korkalainen2019accurate}. Using a single EEG channel further reduces sensing complexity, memory bandwidth, and power consumption, making it well suited for wearable and closed-loop sleep-monitoring applications. Since the primary objective of this work is to develop energy-efficient hardware acceleration rather than optimize the neural network architecture, this representative single-channel model provides an appropriate baseline for evaluating the proposed FPGA accelerator.

From a hardware perspective, the selected model exposes the two dominant workloads targeted in this paper. Its dual CNN branches capture complementary low- and high-frequency sleep features through long- and short-kernel convolutions, while the first 1280-tap, stride-12 convolution alone accounts for more than half of the model’s arithmetic operations. Meanwhile, the BiLSTM introduces recurrent state storage, nonlinear activation functions, and sequential computation that are common challenges among sequence-based sleep-staging models. Consequently, this CNN-BiLSTM network provides a realistic and challenging benchmark for FPGA accelerator development. Moreover, because the proposed FFT-based non-unit-stride convolution, unified hardware architecture, and power-of-two quantization target these fundamental computational patterns rather than a specific network implementation, the proposed techniques are readily applicable to other sleep-staging and EEG models exhibiting similar long-kernel convolution and temporal modeling workloads.
}

\begin{figure}[!t]
    \centering
    \vspace{-1mm}
    \includegraphics[width=1\linewidth]{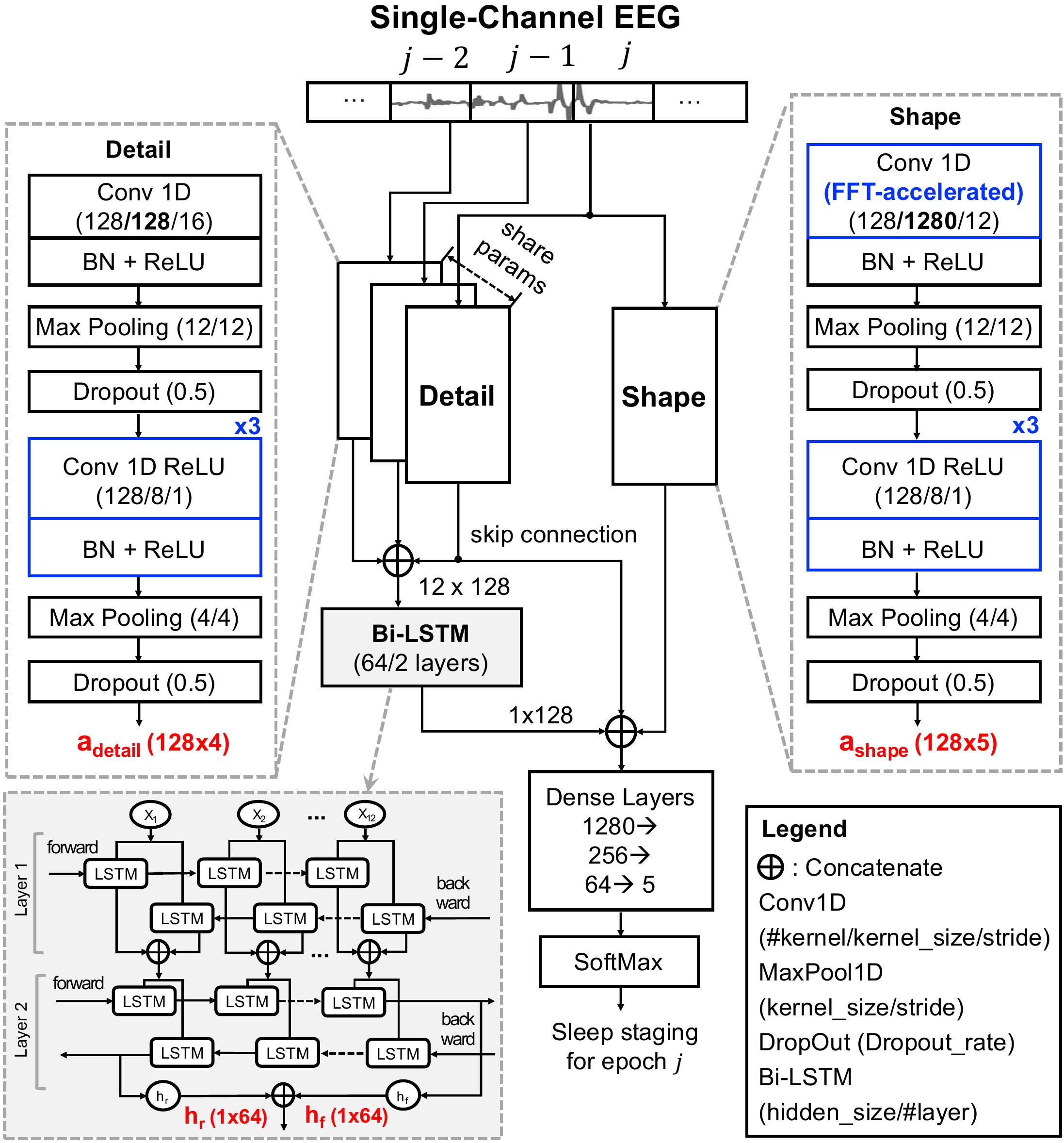}
    \caption{Architecture and key parameters of the CNN-BiLSTM model for sleep staging using a single-channel EEG as input.}
    \label{fig:DL_model}
\end{figure}

{The architecture and key parameters of the CNN-BiLSTM model are shown in Fig.~\ref{fig:DL_model}. The model takes as input 30-second EEG segments sampled at 256~Hz and bandpass filtered between 0.5 and 44~Hz. It consists of two parallel CNN branches followed by a BiLSTM module for temporal sequence modeling.}
The $Shape \ Path$ uses a large kernel of size $1280$ to capture low-frequency features such as Delta waves and K-complexes in the current epoch~\cite{berry2020aasm}, producing a feature tensor $a_{shape}$ of $128 \times 5$. The $Detail \ Path$ processes three consecutive epochs using smaller kernels of size $128$ to extract high-frequency features such as sleep spindles, producing $a_{detail}$ of $128 \times 4$. The BiLSTM module models inter-epoch temporal dependencies across the detail features, and its outputs are concatenated with the $Shape$ and $Detail \ Path$ features and passed through fully connected layers to produce the final five-class sleep stage prediction. {The model was evaluated on the MASS-SS3 dataset~\cite{o2014montreal}, which consists of 62 subjects with sleep-stage annotations following the AASM standard~\cite{berry2020aasm}. The training method, evaluation protocol, and experimental results are presented in Section~IV.
}

\subsection{Fast Convolution Algorithm}

Convolution operations typically dominate the computational load in CNN-based DL models, and this is particularly pronounced in models that employ large-kernel
temporal convolutions for EEG-based inference. In the adopted model, the $Shape\ Path$'s first convolutional layer alone accounts for over 50\% of total model
operations, using kernels of size 1280. Reducing the arithmetic complexity of this layer is therefore the most impactful target for lowering power consumption during hardware inference.

Prior works have employed Winograd and FFT-based fast convolution to accelerate small-kernel CNNs such as AlexNet and VGG-16~\cite{kala2019high, liang2019evaluating}. { However, the large-kernel, non-unit-stride convolutions that dominate our workload fall outside the regime these techniques were designed for, as summarized in Table~\ref{tab:fastconv_comparison}. Winograd is inefficient for large kernels, as partitioning the kernel into small tiles and recombining the results introduces control and memory-movement overhead that does not scale favorably with kernel size.} FFT-based convolution is better suited to larger kernels, but existing accelerators typically rely on small FFT sizes, such as 4- or 8-point transforms, which still incur similar tiling overheads. Large kernels are more appropriately handled using large-point FFT-based convolution, but a challenge arises with non-unit stride. Directly applying a conventional FFT pipeline in this context is highly inefficient, since only a fraction of the computed outputs (e.g., 1 out of every 12 outputs in the model) is actually required.
{ 
To handle non-unit stride, prior fast-convolution methods adopt stride-based decomposition~\cite{yepez2020stride, pan2021accelerate}, in which a non-unit-stride convolution is expressed as a sum of unit-stride sub-convolutions. These methods are formulated around Winograd and accumulate the sub-convolution results in the time domain. A direct combination of stride decomposition with FFT would instead transform each of the $S$ unit-stride sub-convolutions independently and accumulate them in the time domain, but this requires a separate IFFT per sub-convolution and therefore scales as $O(SN\log N)$, eroding much of the benefit of operating in the frequency domain.
}

\begin{table}[!ht]
\centering
\caption{Comparison of the Proposed Fast Convolution with Prior Strided / FFT-Based Approaches}
\label{tab:fastconv_comparison}
\renewcommand{\arraystretch}{1}
\setlength{\tabcolsep}{1.5pt}
\begin{threeparttable}
\begin{tabular}{@{}>{\raggedright\arraybackslash}p{2.6cm}ccccc@{}}
\toprule
\textbf{Method} &
\textbf{\shortstack{Kernel\\Regime}} &
\textbf{\shortstack{Stride\\Handling}} &
\textbf{\shortstack{Post-Proc.\\Complexity\\For FFT}} &
\textbf{\shortstack{IFFTs/\\Block}} &
\textbf{\shortstack{Engine\\Reuse}} \\
\midrule
Winograd fast conv.~\cite{kala2019high}&
Small & \shortstack{None\\(discards)} & --- & --- & No \\
 Conventional FFT fast conv.~\cite{liang2021fcnnlib}& \shortstack{Large}& \shortstack{None\\(discards)} & $O(N\log N)$* & 1&No \\
Winograd stride decomp.~\cite{yepez2020stride, pan2021accelerate}&
Small & Decomp. & --- & --- & No \\
 stride decomp.~\cite{yepez2020stride, pan2021accelerate} + FFT& Large& Decomp. & $O(SN\log N)$& $S$&---\\
\textbf{This work} &
\textbf{\shortstack{Large\\(1280)}} & \textbf{Decomp.} &
${O(N\log N)}$ & \textbf{1} & \textbf{Yes} \\
\bottomrule
\end{tabular}
\begin{tablenotes} 
\item {$^{*}$Computes the full-resolution output; under non-unit stride the majority of outputs are discarded, so useful-work efficiency is far lower than the complexity figure suggests.}
\end{tablenotes}
\end{threeparttable}
\end{table}

We introduce a new formulation method building on prior stride-based decomposition approaches~\cite{yepez2020stride, pan2021accelerate}. { In contrast to the time-domain accumulation used by these prior methods, our formulation performs the intermediate accumulation directly in the frequency domain, exploiting the linearity of the Fourier transform so that a sum in the time domain is preserved as a sum in the frequency domain. This allows the IFFT to be deferred until after all sub-convolutions have been summed, reducing post-processing complexity from $O(SN\log N)$, which would otherwise require $S$ IFFTs, to $O(N\log N)$ with a single IFFT per block. To the best of our knowledge, this work is the first to apply large-point FFT convolution to large-kernel layers while explicitly addressing the inefficiency introduced by non-unit strides.}

We implement this approach using an $N$-point radix-2 FFT to perform the necessary Discrete Fourier Transform (DFT) and Inverse DFT (IDFT) operations. The sampling decomposition algorithm is detailed in Algorithm~\ref{algo:decompose}, which takes a 1-D array (input or weights) and returns a matrix where every $s$-th input is placed in a different row, enabling efficient processing.
\begin{algorithm}[!h]
\caption{Decompose(A, s)}
\begin{algorithmic}
\STATE \textbf{Input:} 1-D array $A$ of $n$ and integer $s$ representing parameter of stride
\STATE \textbf{Output:} Decomposed matrix $s$ by number of points for each stride
\STATE $remainder = n \: mod\: s$
\STATE $size = ceil(n\: /\: s)$
\STATE $output = [s; \: size]$
\FOR{$i = 0;\: i < s$}
\IF{$i < remainder$}
\FOR{$j = 0;\: j < size$}
\STATE $ output[i,j] = A[j \cdot s+i] $
\ENDFOR
\ELSE
\FOR{$j = 0;\: j < size - 1$}
\STATE $ output[i,j] = A[j \cdot s+i] $
\ENDFOR
\ENDIF
\ENDFOR
\RETURN $output$
\end{algorithmic}
\label{algo:decompose}
\end{algorithm}

The result of a non-unit stride convolution can be obtained by performing unit-stride convolutions on the decomposed input and kernel matrices. The detailed procedure is described in Algorithm~\ref{algo:decomposed_conv}. Specifically, for each stride index, the non-unit stride convolution is computed as the element-wise sum of the unit-stride convolution outputs corresponding to that index across the decomposed matrices. {This structure can be exploited from a hardware perspective. Given the linearity of the Fourier transform, addition in the time domain is preserved as addition in the frequency domain. Consequently, the decomposed convolution scheme can be optimized by computing the unit‑stride convolution using the FFT and performing the intermediate accumulations directly in the frequency domain. Only a single IFFT is required for each block, reducing the arithmetic complexity from $O(SN\log(N))$ to $O(N\log N)$. Notably, this accumulation pattern resembles a convolution itself, a property that we also leverage in our hardware design.}

\begin{algorithm}[!ht]
\caption{Decompose\_convolution(A, s)}
\begin{algorithmic}
\STATE \textbf{Input:} 2-D input matrix $A$ of size ($s$, $in\_size$), 2-D kernel matrix W of size ($s$, $kernel\_size$),  stride $s$
\STATE \textbf{Output:} convolution result of stride $s$
\STATE Let conv\_1d(a, b) be a function that computes the unit stride convolution of two inputs a and b
\STATE $output = [in\_size - kernel\_size + 1]$
\STATE $output$ is initialized with each element of a value of 0
\FOR{$i = 0;\: i < s$}
\STATE $ result[i] \leftarrow conv\_1d(A[i],W[i]) $
\FOR{$ j = 0;\: j < in\_size - kernel\_size + 1$}
\STATE $output[j] \leftarrow output[j] + result[j]$
\ENDFOR
\ENDFOR
\RETURN $output$
\end{algorithmic}
\label{algo:decomposed_conv}
\end{algorithm}

An $N$-point FFT, where $N$ is a power-of-two number, is well-suited for hardware implementation due to its butterfly computation structure, which enables efficient in-place execution. Similarly, the Inverse FFT (IFFT), an efficient method for computing the IDFT, can be implemented using the same FFT hardware. One common approach involves reordering the frequency-domain elements and scaling the result by $1/N$, since $IDFT(X[k]) = DFT(\frac{1}{N}X[((-k))_N])$. This allows the FFT and IFFT to share control logic and hardware resources, improving overall reuse.

Several additional optimizations are applied to further reduce computational cost. Since convolutional kernels are fixed across inferences, their frequency-domain representations can be precomputed and stored as model parameters. Decomposed kernels are time reversed and zero-padded to length $N$. When the kernel size divided by stride has a non-zero remainder, smaller decomposed kernels receive an additional leading zero to align with the outputs of longer kernels. The storage overhead introduced by precomputed
weights is mitigated by conjugate symmetry in the frequency domain. Since the weights are real-valued in the time domain, only the first $(N / 2) + 1$ of the total $N$ frequency components need to be stored, reducing the weight storage by nearly half. The same symmetry also reduces the frequency-domain multiplication to $(N/2)+1$ complex operations per block.

Finally, since the output of the convolution layers is real-valued, the IFFT computations can be halved using pairwise processing~\cite{matusiak2001implementing}. Given two frequency-domain sequences $X_1[k]$ and $X_2[k]$ corresponding to real-valued time-domain signals, a composite sequence $X_3[k] = X_1[k] + jX_2[k]$ is formed. After applying the IFFT, the real part of $IDFT(X_3[k])$ yields $x_1[n]$, while the imaginary part yields $x_2[n]$. This technique reduces the number of IFFT computations by approximately half.

\subsection{PoT Quantization of the Model}

Quantization reduces the computational and memory footprint of DL models by mapping floating-point parameters to fixed-point representations.
We employ PoT quantization, where the scale factor mapping the floating-point range to the fixed-point grid is constrained to a power of two, enabling all scaling operations
to be implemented as bit-shifts and eliminating the need for multipliers~\cite{przewlocka2022power,nagel2021white}.
FFT-based pipelines impose an additional challenge for quantization. Specifically, the bit-width must remain large enough to maintain numerical stability across butterfly stages, where rounding errors accumulate with each successive computation~\cite{welch2003fixed}. To balance numerical fidelity against hardware cost, we evaluated both 8-bit and 16-bit PoT quantization (to be discussed in Section IV-B).

Under the PoT quantization scheme, the representation $range$ is directly related to the factor $scale$ through $range \approx scale \times 2^{bitwidth} $. The range of biases is generally larger than that of individual activations and weights. Prior work uses 32-bit representations for biases~\cite{nagel2021white,jacob2018quantization}.
To simplify hardware memory management, we use a uniform bit-width for activations, weights, and biases throughout inference, assigning a separate scale to each component, making the scaling factor a key optimization parameter per kernel and activation.

\begin{figure}[!ht]
    \centering
    \includegraphics[width=\columnwidth]{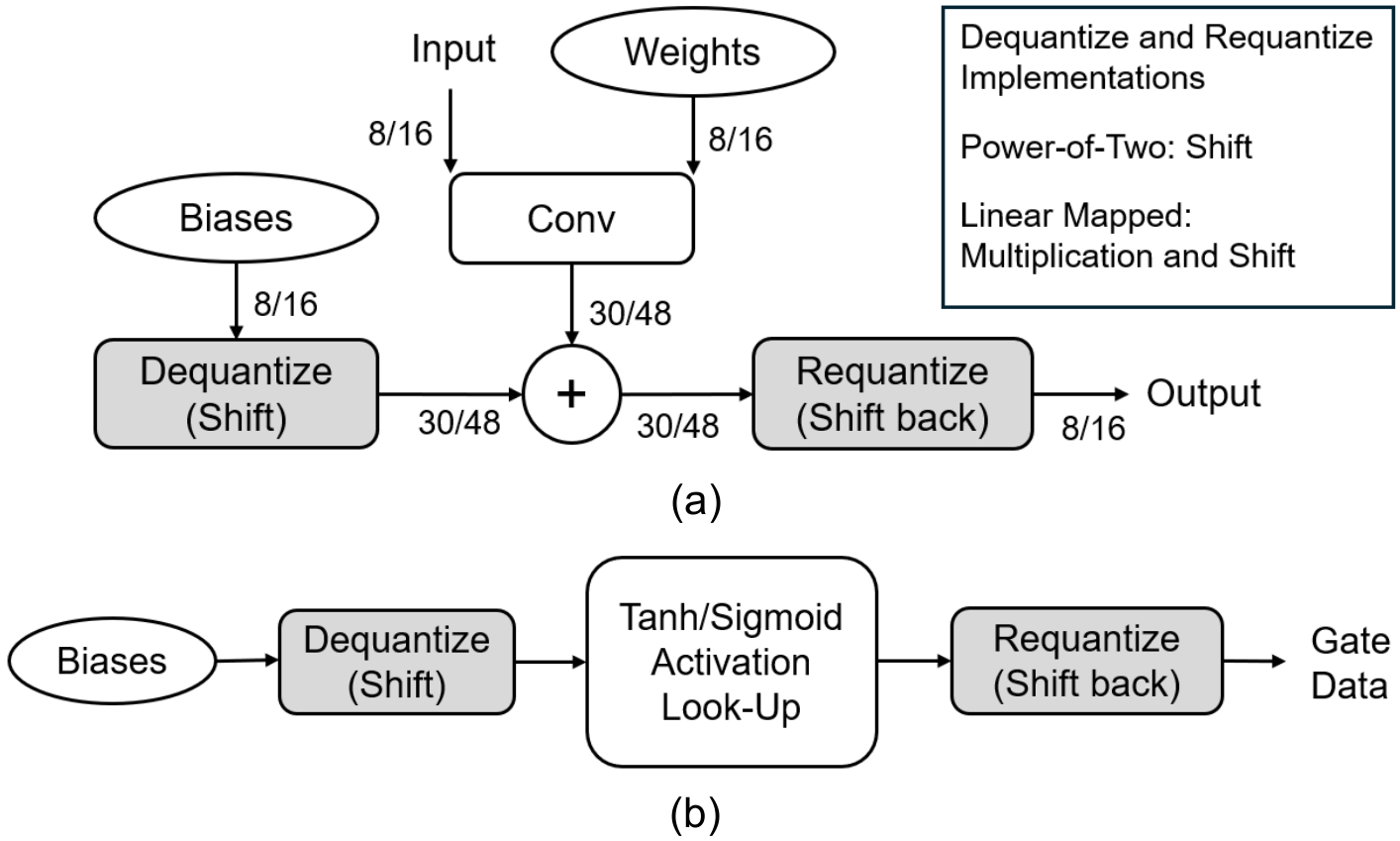} \vspace{-6mm}
    \caption{(a) Quantized convolution / dense operation flow specific to our hardware using PoT quantizer, and (b) quantized \texttt{tanh} or \texttt{sigmoid} computation operation using interpolation of LUT entries.}
    \label{fig:quant_scheme}
\end{figure}

The computational flow for convolution and dense operations is shown in Fig.~\ref{fig:quant_scheme}(a).
At the start of each convolution, we first add the de-quantized bias to ensure that its scale matches that of the product of the activation and weight. After adding the bias, the activation is multiplied with the weights, and the result is added to the sum. In the adopted model, the output activations from a convolution operation can grow as large as 42 bits for 16-bit inference and 26 bits for 8-bit inference. To maintain a consistent bit width for activations throughout inference, the outputs need to be re-quantized back to 8 or 16 bits.

The BiLSTM module follows the same quantized data flow, but additionally requires quantized evaluation of non-linear activation functions, \texttt{tanh} and \texttt{sigmoid}. These are computed via linear interpolation over a LUT, as detailed in Section~III-C. The computation flow, illustrated in Fig.~\ref{fig:quant_scheme}(b), performs interpolation using the same scale as the inputs stored in the LUT.

For FFT-based convolution, quantization is applied across the transform pipeline. Two strategies are considered~\cite{gupta2016accurate}. Strategy~1 preserves all intermediate bits across FFT butterfly stages, allowing bit-width to grow to maintain accuracy. Strategy~2 truncates results to a fixed 16-bit width after each stage, reducing memory usage at the cost of increased quantization noise. In both strategies, the final activation is scaled back to 16 bits before being written to main memory. A comparison of signal-to-quantization-noise ratio (SQNR) and hardware requirements for both strategies is provided in Table~\ref{table:FFT_quant_perf}. Strategy~1 is selected: the additional 0.5~KB of memory it requires is negligible relative to total CNN weight storage, and its quantization noise is concentrated in the early FFT stages~\cite{gupta2016accurate}, making it well-matched to the DIT FFT implementation used in this work, where the first butterfly stage requires no twiddle factor multiplication.

\begin{table}[!ht]
\centering
\caption{Comparison Between Quantization Strategies}
\label{table:FFT_quant_perf}
\renewcommand{\arraystretch}{1.2}
\begin{tabular}{ccc}
\hline
\makecell{FFT Stage-wise\\ Bit Precisions} & \makecell{Quantization \\Strategy 1~\cite{gupta2016accurate}} & \makecell{Quantization \\Strategy 2~\cite{gupta2016accurate}} \\
\hline
Stage 1 (bits)  & 17 & 16 \\
Stage 2 (bits)  & 18 & 16 \\
Stage 3 (bits)  & 19 & 16 \\
Stage 4 (bits)  & 20 & 16 \\
Stage 5 (bits)  & 21 & 16 \\
Stage 6 (bits)  & 22 & 16 \\
Stage 7 (bits)  & 23 & 16 \\
Stage 8 (bits)  & 24 & 16 \\
\hline
SQNR Theoretical (dB) & 97.8~\cite{gupta2016accurate} & 63.1~\cite{gupta2016accurate} \\
SQNR Simulated (dB)   & 98.1~\cite{gupta2016accurate} & 63.6~\cite{gupta2016accurate} \\
RAM Size (KB)          & 1.5 & 1.0 \\
\hline
\end{tabular}
\end{table}

\begin{figure*}[!ht]
    \centering
    \includegraphics[width=0.93\textwidth]{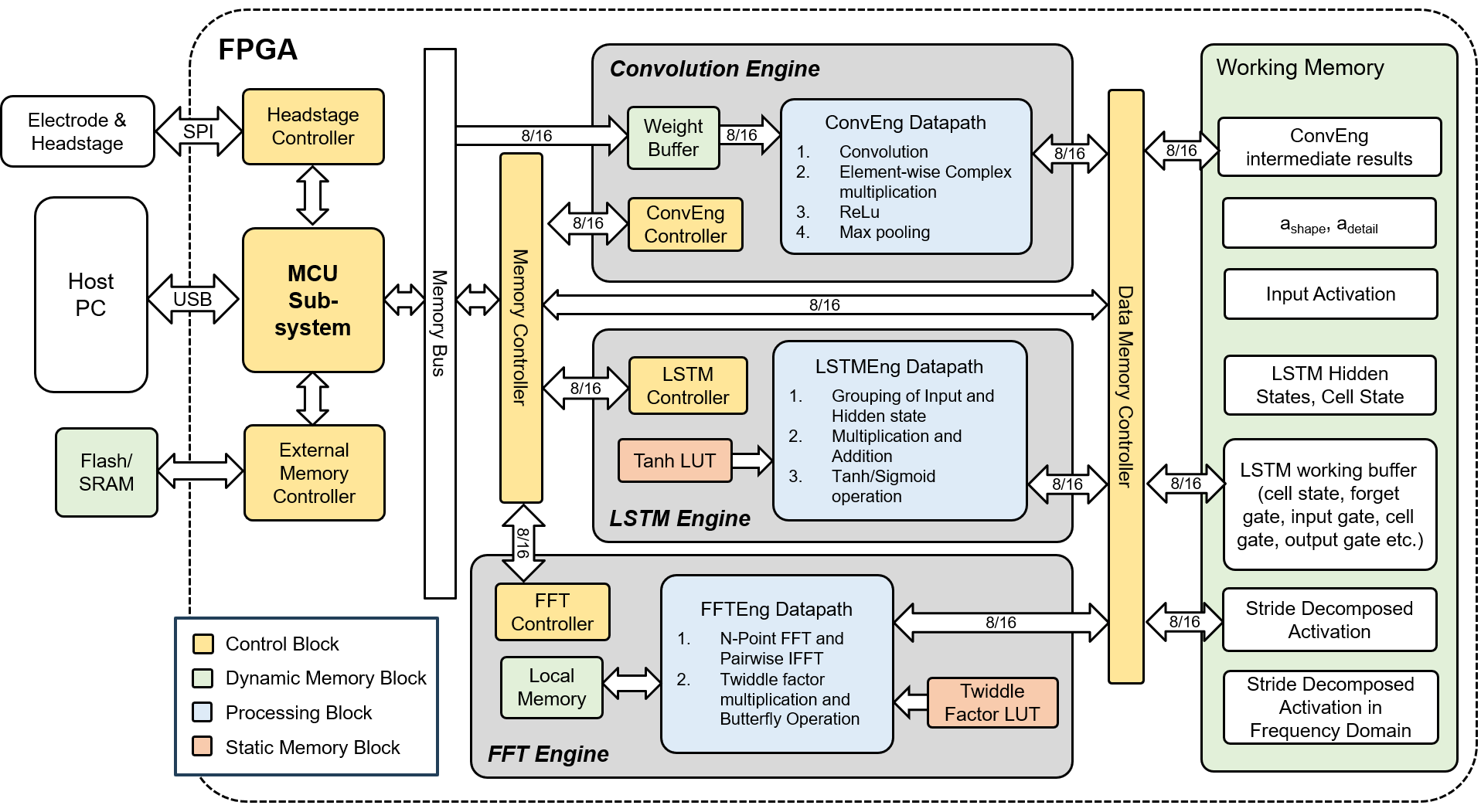}
    \caption{Top-level hardware system architecture of the proposed FPGA accelerator. The design consists of an MCU for system control, convolution engine, BiLSTM engine, and data memory for weights and intermediate results.}
    \label{fig:sys_architecture}
\end{figure*}

\section{FPGA Acceleration of the DL Model}

\subsection{Overview of the FPGA-based Accelerator}

Fig.~\ref{fig:sys_architecture} shows the block diagram of the developed FPGA accelerator. The accelerator is implemented using a combination of custom RTL and vendor IPs. Compute engines (convolution, FFT, and BiLSTM) are implemented in custom RTL for precise datapath control. Standard AXI interfaces and memory controllers are integrated via vendor IP to facilitate communication with a microcontroller (MCU) sub-system. Model weights are stored in external flash or on-chip SRAM and are streamed into the programmable logic as required.

The MCU manages system control tasks. Through memory-mapped configuration registers, the MCU configures each engine, schedules the execution of each layer, and coordinates the movement of data between shared activation buffers. This MCU-based control structure enables flexible sequencing of operations and efficient reuse of hardware resources across the convolution, FFT, and BiLSTM computations. Its programmability also provides a convenient path to adapt the hardware to future changes
in the DL model architecture.

The FFT engine supports $N$-point DIT in-place FFT and IFFT, sharing the same data flow for both operations with an additional pre-processing stage for the IFFT. The convolution engine processes the EEG signal from the MCU subsystem and is responsible for performing convolutions, dense layer operations, and element-wise multiplications. This includes the computation of $a_{shape}$, $a_{detail}$, the dense layers, and the complex element-wise multiplications required for time domain convolution in the frequency domain. The BiLSTM engine reuses the convolution engine to compute the forward and backward hidden states ($h_{f}$ and $h_{r}$) for the BiLSTM operations on $a_{detail}$. Model weights in the programmable logic are stored in a dedicated weight buffer and are accessible only by the convolution engine, while intermediate activations are stored in working memory.

\subsection{Acceleration of the CNN}

Prior CNN accelerators have exploited fast convolution techniques, including radix-2 FFT or Winograd~\cite{liang2019evaluating, liang2021fcnnlib, yang2023flexible}. Additionally, past work explored stride-based decomposition methods to optimize convolution operations~\cite{yepez2020stride}. Other works have leveraged model attributes like sparsity to reduce computational load and minimize off-chip memory transactions~\cite{lu2019efficient, wu2024edge}. However, these approaches often depend on complex model structures that are not well-suited for wearable applications that operate under tight power and hardware constraints.

Our CNN accelerator is designed to optimize both computational efficiency and energy consumption. In addition to the fast convolution algorithm described in Section II, power consumption is further reduced by minimizing the number of off-chip memory accesses, such as flash memory~\cite{sze2017efficient}. Rather than relying on sparse-weight detection logic, the design adopts a dense, regular architecture that ensures low hardware overhead and consistent latency, both of which are critical for real-time sleep staging. To this end, three complementary memory and scheduling optimizations are introduced:

\begin{itemize}
    \item \textbf{Block convolution:}
        We divide large convolution layers into smaller blocks using overlap-save block convolution~\cite{oppenheim1999discrete}. This approach enables fixed $N$-point FFT-based fast convolution across varying kernel sizes, while also reducing peak memory usage during inference for both conventional and fast convolution paths.
    \item \textbf{Depth-first execution:}
        We integrate a depth-first execution strategy for both standard convolutions and non-unit stride FFT-based convolutions. This scheduling aligns with the model topology and enables efficient reuse of hardware resources. By executing the decomposition algorithm depth-first, we reduce the additional memory cost from $O(NS)$ to $O(N)$, where $N$ is the FFT size and $S$ is the stride. This substantially lowers intermediate activation storage and minimizes additional hardware complexity.
    \item \textbf{Concurrent engine execution:}
        We design separate convolution and FFT engines. To exploit the convolution-like structure of the decomposition algorithm, we extend the convolution engine to support complex-valued convolution. This enables the decomposition algorithm's post-processing to run in parallel with the IFFT operations, reducing the additional hardware complexity required to support the decomposition algorithm. Building on this capability, we introduce a custom scheduling strategy that minimizes latency by overlapping FFT, point-wise multiplication, and IFFT operations across blocks.
\end{itemize}

\begin{figure}[!b]
    \centering
    \includegraphics[width=0.43\textwidth]{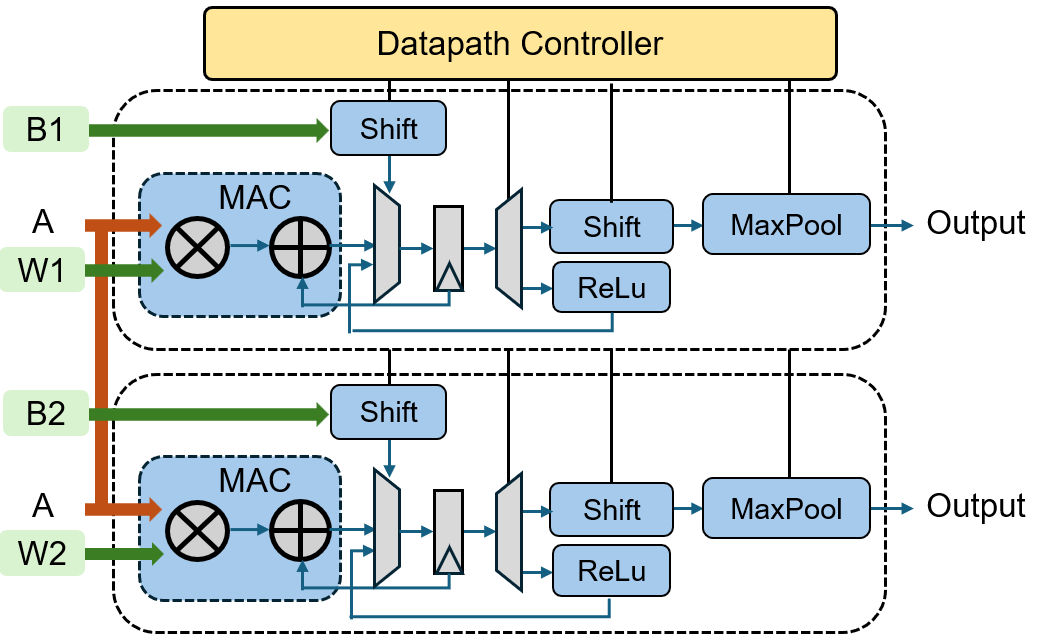}
    \caption{The convolution computation data path for two PEs. The data flows through MAC, then ReLU is applied and shifted back to the incoming bit width. MaxPool is the final stage before the result is written to data memory.}
    \label{fig:Conv_Comp}
\end{figure}

The convolution engine supports two core arithmetic operations: conventional convolution and element-wise complex multiplication. The engine is managed by a controller that allocates a set of processing elements (PEs) according to the configuration registers set by the MCU. For conventional convolution, the engine uses a depth-first execution data path to reduce on-chip memory usage. The data flow for two PEs is illustrated in Fig.~\ref{fig:Conv_Comp}.
During convolution, the bias is first loaded into partial sum registers and scaled according to the precision of the input. Then, weights and activations are multiplied and accumulated. The partial sum registers are 48-bit for 16-bit inference and 30-bit wide for 8-bit inference to support model flexibility. Once the multiply-and-accumulate (MAC) stage is complete, the output is processed locally through ReLU activation and re-quantized using a shift-back operation. Finally, max-pooling is applied before the result is written back to the data memory.

The controller is configurable via control registers that enable ReLU operations and set the max-pooling size, supporting different layers and potential future model changes. Applying ReLU and max-pooling locally before memory write-back reduces both memory accesses and memory footprint. Multiple PEs can be instantiated based on the latency requirements, and all PEs share a common activation memory.

To improve resource sharing between conventional convolution operations and FFT fast convolution operations, element-wise multiplications are efficiently handled by the same PEs with minimal overhead. The convolution engine is further optimized by unifying the control logic for standard convolution and the MAC operations required by the decomposition algorithm. As discussed in Section~II-C, the decomposition post-processing exhibits a convolution-like structure, enabling both functional reuse and a reduction of its memory complexity from $O(NS)$ to $O(N)$. This allows the control logic, weight memory, and computation units of the convolution engine to be directly reused for the decomposition post-processing, enabling concurrent engine execution with minimal additional hardware cost while reducing intermediate activation storage.

\begin{figure}[!ht]
    \centering
    \includegraphics[width=\columnwidth]{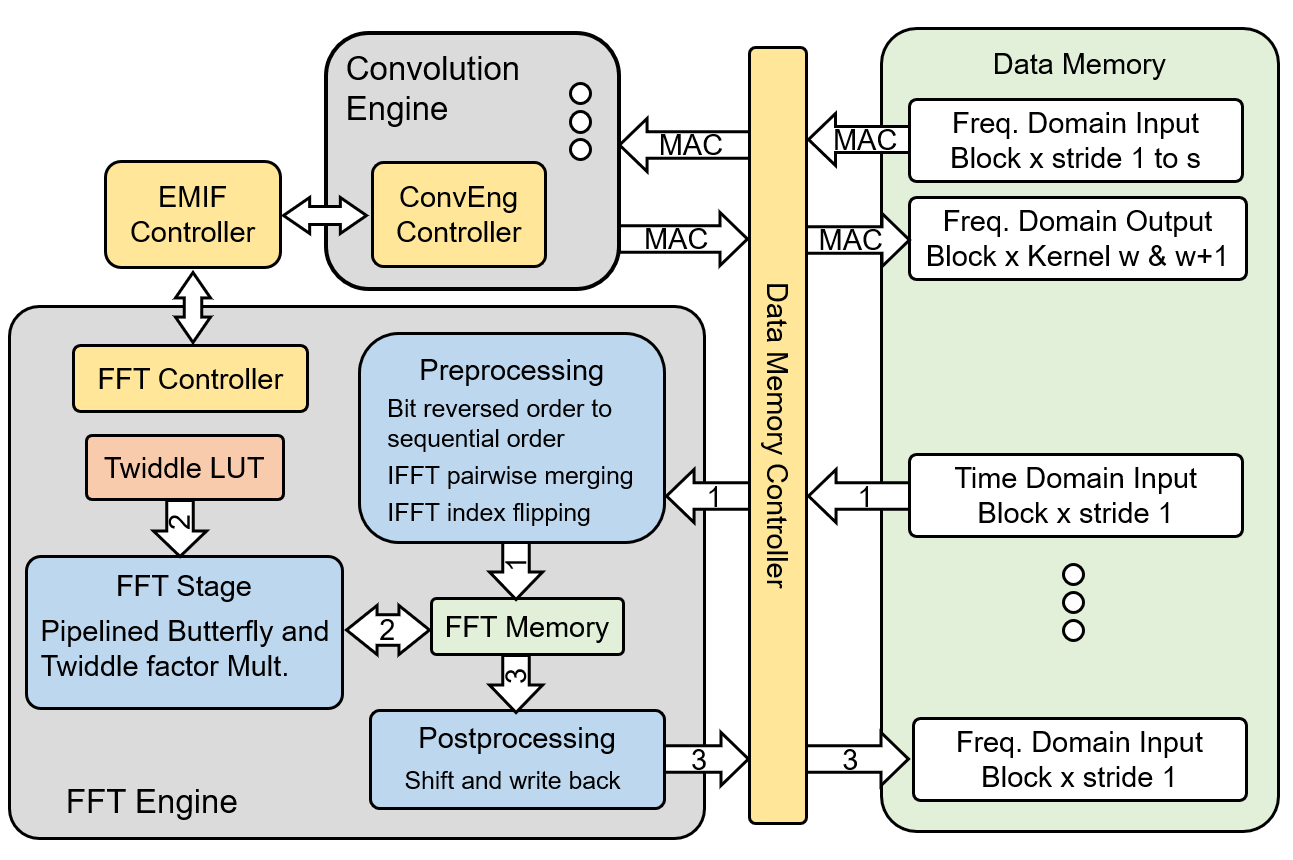}
    \caption{Block diagram of the FFT processing engine. The sequence of data memory access by each of the modules are labeled in the arrows. The controller block is connected to all components to manage states. It can be configured to control both the FFT engine and convolution engine to run concurrently for the complex MAC and IFFT, respectively.}
    \label{fig:FFT_PE}
\end{figure}

The FFT engine includes dedicated LUTs for the real and imaginary components of the twiddle factors. As illustrated in Fig.~\ref{fig:FFT_PE}, a memory block stores the intermediate results of the FFT stages, with each memory word being 24-bit wide. For each FFT stage, we leverage the in-place computation characteristic of the DIT FFT. Twiddle factor multiplication and butterfly operations are fused into a single step, allowing both to be completed before writing the results back to the FFT memory. This data flow is memory-bound, requiring four reads and four writes per butterfly operation. A single multiplier is used, processing each pair of elements in four cycles to meet
throughput targets.

To perform an FFT operation, the engine first pre-processes the input by reading data memory sequentially and writing each element into the FFT memory in bit-reversed order. It then computes each FFT stage accordingly. After the final FFT stage, the controller applies a shift-back operation to scale the results according to the quantization factor, ensuring the output retains the same bit-width as the input.

Similarly, IFFT is computed using the same FFT engine. As noted in Section II, computing the IFFT for time-domain real-valued signals in pairs reduces computational overhead. Accordingly, the IFFT pre-processing combines the two sequences and reverses their order before writing into the bit-reversed indices. A key difference is that the shift back value is configured to increase by $\log_2N$, since FFT and IFFT are symmetric but scaled differently by $1/N$.

The hardware uses a fixed $N$-point FFT. To accommodate varying layer sizes, an overlap-save block convolution strategy is employed. This avoids the overhead during inference from storing and accumulating overlapping regions required by the overlap-add method. Given CNN kernel length $K$ and stride $S$, based on our decomposition algorithm, consecutive blocks overlap by $(\lceil K / S \rceil-1)$ samples and advance by $H=N-(\lceil K / S \rceil-1)$. Post IFFT, the first $(\lceil K / S \rceil-1)$ samples are discarded and the remaining $N-\lceil K / S \rceil+1$ samples are the valid linear convolution outputs.

Block convolution also mitigates the peak memory usage of large kernel layers, effectively reducing on-chip memory requirements. For large kernels, storing all associated weights on-chip before computation requires a substantial amount of memory. The convolution layer can be divided along the channel dimension to resemble the overlap-save method. This allows each output channel of the convolution layer to be computed independently. As a result, at most 1280 weights per PE need to be stored on-chip at any given time.

\begin{figure}[!ht]
    \centering
    \includegraphics[width=0.48\textwidth]{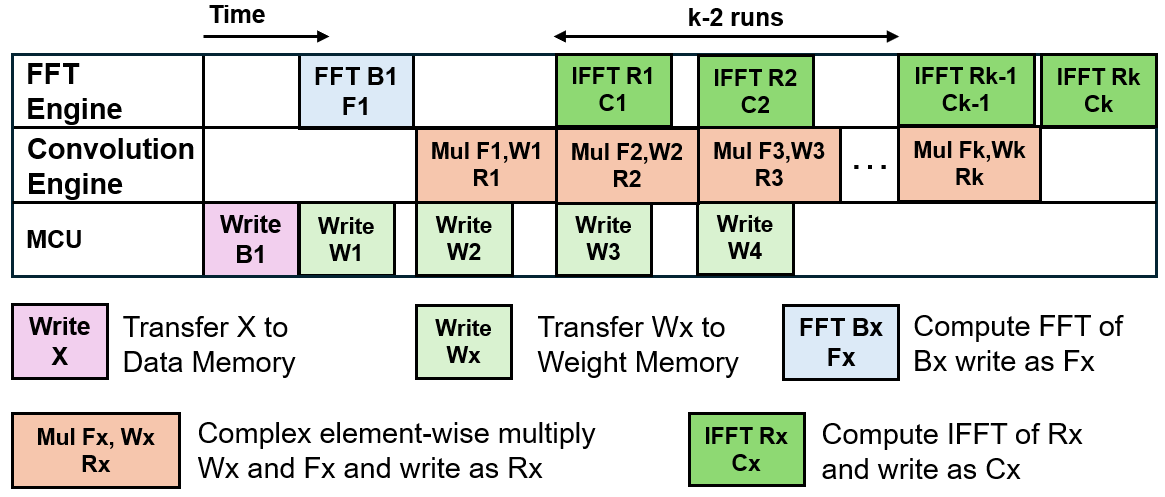}
    \caption{Execution trace of the convolution and FFT engine showing proposed task scheduling and concurrency for fast convolution with $k$ kernel sets. After initial data transfer and FFT operations, IFFT, element-wise complex multiplication and weight transfer run concurrently. The timing is illustrative and not to scale. }
    \label{fig:FFT_schedule}
\end{figure}

Concurrent execution of the convolution and FFT engines is illustrated in Fig.~\ref{fig:FFT_schedule}. After the initial FFT and complex MAC operations, IFFT, MAC, and weight transfer can run in parallel until the second-to-last kernel set IFFT operation. This scheduling strategy enables the FFT and convolution engines to operate in parallel for the majority of each input block’s processing time, improving overall hardware utilization and reducing latency.

\subsection{Acceleration of the BiLSTM Network}

Prior LSTM accelerators have explored gate-level optimizations, memory-aware scheduling between DRAM and BRAM~\cite{guan2017fpga}, matrix multiplication tiling, and binary adder trees to enhance parallelism~\cite{zhang2017implementation}. More recently, the CORDIC algorithm has been proposed for low-power computation of non-linear activation functions~\cite{mohamed2023unified}. Despite these advances, most designs target general-purpose or high-performance applications and do not fully address the constraints of biomedical workloads, where low power, compact area, and system-level integration are essential. Low-power LSTM acceleration for biomedical inference therefore remains largely unexplored.

In this work, the BiLSTM processing engine reuses the existing convolution engine for all major dense computations, reducing hardware redundancy. Non-linear activation functions are efficiently computed using a linear interpolation LUT, which offers a compact alternative to more complex arithmetic units. Furthermore, BiLSTM operations are partitioned into forward and backward passes, enabling tiling strategies that reduce memory footprint and simplify control logic.

\begin{figure}[!ht]
    \centering
    \includegraphics[width=0.48\textwidth]{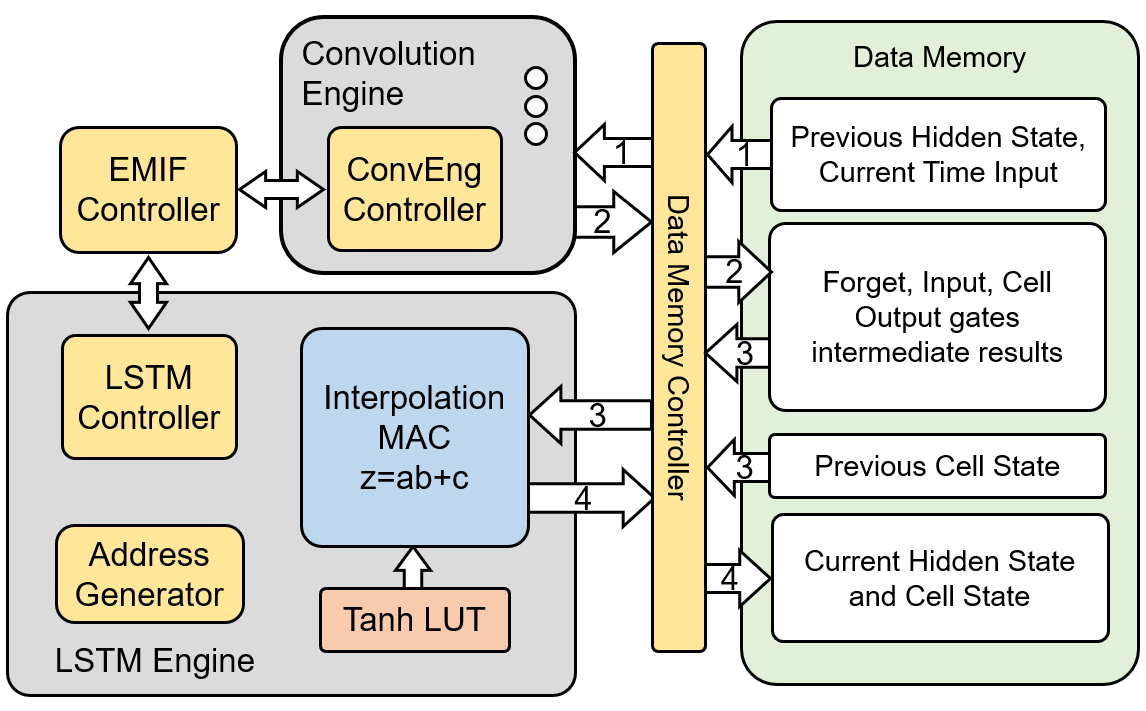}
    \caption{Block diagram of the BiLSTM processing engine. The sequence of data memory accesses through the data memory controller by each of the modules are labeled. The connections between controller blocks are left blank as they are used throughout the computation.}
    \label{fig:LSTM_block}
\end{figure}

The block diagram of the BiLSTM engine is shown in Fig.~\ref{fig:LSTM_block}. The engine interfaces with the External Memory Interface (EMIF) controller to configure and control the convolution engine, which is repurposed for computing the dense operations on both input and hidden state vectors. An address generator within the BiLSTM engine manages memory addresses for the hidden state, cell state, and working memory. The LUT for \texttt{tanh} also supports \texttt{sigmoid}, enabling efficient computation of non-linear activation functions. To perform a BiLSTM operation, the engine first concatenates the current sequential input and the hidden state by reading from and writing to the working memory. It then delegates the dense computations for the forget gate, input gate, cell gate, and output gate to the convolution engine. Afterward, the BiLSTM engine performs the necessary \texttt{sigmoid} and \texttt{tanh} operations, along with multiplications and additions, to update the cell and hidden states. For flexibility, the BiLSTM engine operates in a designated memory region within the working memory to store the outputs of each gate, as well as the hidden and cell states for each time step.

Reusing the convolution engine for compute-intensive operations allows us to optimize area efficiency. However, the BiLSTM requires a large set of weights for each gate. In the flow of an RNN, these weights are reused for each sequential time step. To prevent unnecessary rewriting of weights, all gate weights are stored locally in hardware. Therefore, the size of the weight buffer is required to be large enough to hold the complete set of BiLSTM gate weights at any given time.

Another complexity for BiLSTM inference on hardware is the computation of the two non-linear activation functions, \texttt{tanh} and \texttt{sigmoid}. To approximate \texttt{tanh} under quantization, an interpolation method is developed using a 38-entry, 12-bit LUT. To account for the larger derivative of \texttt{tanh} near zero, the data points are distributed non-uniformly, with 16 entries spanning the range from 0 to 1. Leveraging the symmetry of \texttt{tanh}, only the positive half of the curve is stored. The interpolation algorithm shares the multiplier used by other BiLSTM operations, minimizing hardware resource usage.

To reuse the data points in the \texttt{tanh} table for computing the \texttt{sigmoid} activation, we can manipulate the \texttt{sigmoid} formula into a hardware-friendly form based on \texttt{tanh}:

\begin{equation}\label{sigmoid_function}
\begin{aligned}
    \sigma(x) &= \frac{1}{1 + e^{-x}} = \frac{e^{x/2}}{e^{x/2} + e^{-x/2}} \\
              &= \frac{1}{2} (1+\frac{e^{x/2} - e^{-x/2}}{e^{x/2} + e^{-x/2}}) \\
              &= \frac{1}{2} (1 + \tanh(\frac{x}{2}))
\end{aligned}
\end{equation}

\begin{figure}[!ht]
    \centering
    \includegraphics[width=\columnwidth]{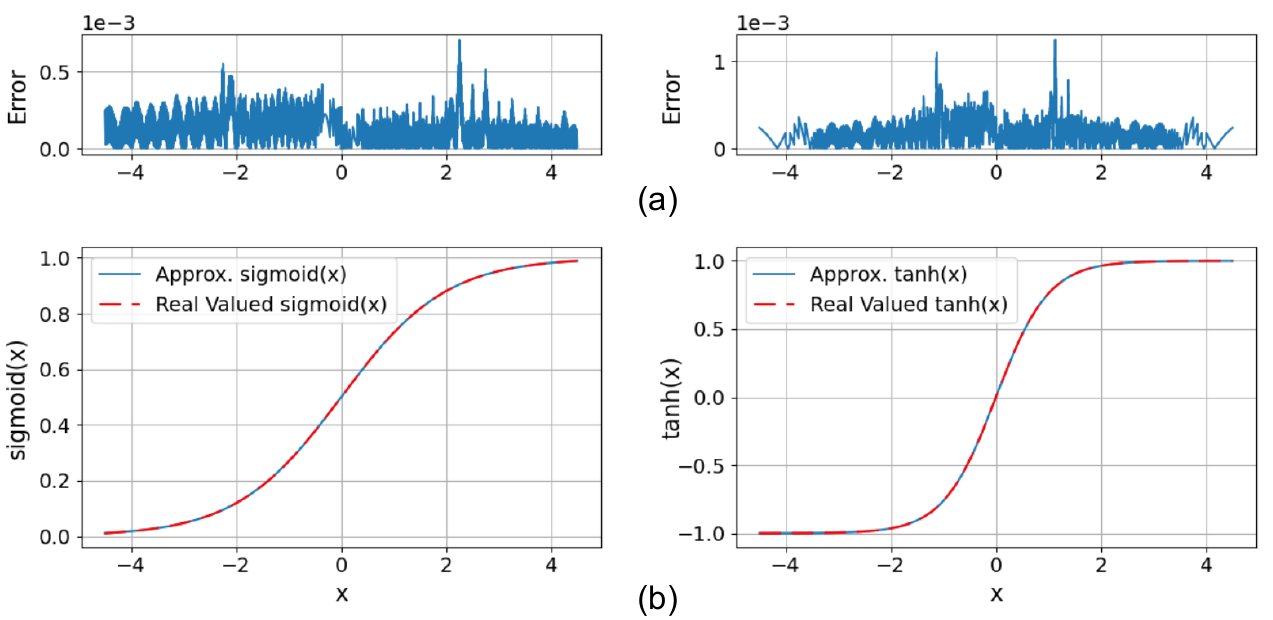}
    \caption{(a) The approximation error between the ideal \texttt{sigmoid} and \texttt{tanh} activation functions and their hardware approximations. (b) The overlay of the ideal and approximation functions.}
    \label{fig:tanh_approx}
\end{figure}

As shown in Eq.~\eqref{sigmoid_function}, once the \texttt{tanh} value is computed, a single shift operation yields the corresponding \texttt{sigmoid} result. This approach allows both functions to be computed using a single LUT. If higher precision is required for either function, increasing the number of LUT entries improves the accuracy of both functions. Fig.~\ref{fig:tanh_approx}(a) shows the approximation error between the ideal \texttt{tanh} and \texttt{sigmoid} activation functions and their corresponding hardware approximations, and Fig.~\ref{fig:tanh_approx}(b) presents an overlay of the ideal and approximated curves. The results confirm that both approximations closely match their real-valued counterparts with minimal error across the input range.

\section{Experimental Results}

{
This section experimentally evaluates the proposed algorithm-hardware co-design from three complementary perspectives. First, the computational benefits of the proposed FFT-based non-unit stride convolution are quantified through arithmetic complexity and memory access. Next, the impact of the developed PoT quantization scheme is evaluated in terms of signal fidelity and sleep staging accuracy relative to the floating-point baseline. Finally, FPGA implementation and testing results, including resource utilization, power consumption, and measured inference latency, are presented and analyzed. 

}

\subsection{FFT Testing Results}

\begin{figure}[!ht]
    \centering
    \includegraphics[width=0.48\textwidth]{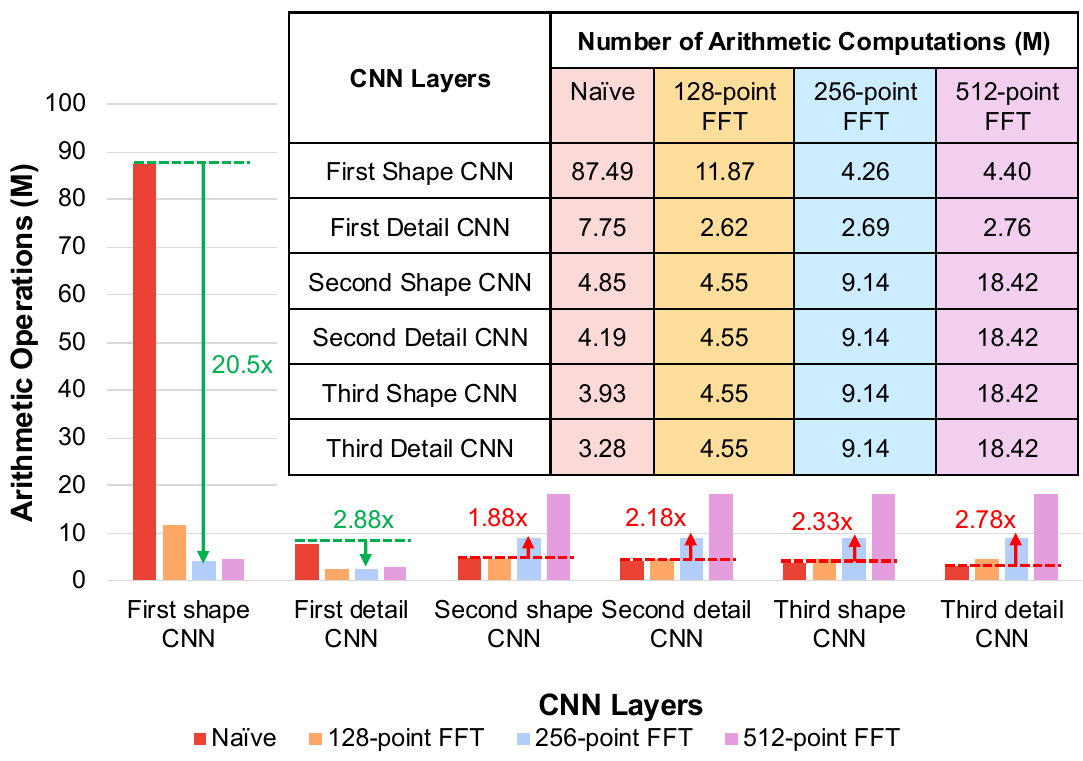}
    \caption{Comparison of real arithmetic operations without FFT and using different $N$-point FFT fast convolution. Using a 256-point FFT, a 20.5$\times$ reduction is achieved in the $a_{shape}$ layer computation.}
    \label{fig:FFT_sweep}
\end{figure}

The computation reduction enabled by the proposed stride-based decomposition algorithm is evaluated in this subsection. To determine the optimal implementation for an $N$-point DIT FFT, we evaluated different values of $N$, where all decomposed kernels in the model fit within the $N$-point FFT buffer, for their suitability in computing $a_{shape}$. Fig.~\ref{fig:FFT_sweep} shows the number of computations required when 128-point, 256-point, and 512-point DIT FFTs are applied to the first three CNN layers of $a_{shape}$ and $a_{detail}$ in the DL model.

For layers with long kernels and long inputs, such as the first layer of $a_{shape}$, FFT provides substantial benefits: a 256-point or 512-point FFT reduces the number of real-valued operations by approximately 20.5$\times$. However, this benefit does not generalize across all layers. In later layers with short kernels (e.g., 8-point kernels in the second and third CNN layers), the zero-padding required to map the computation onto a large-$N$ FFT eliminates any advantage, making much of the FFT computation redundant. The fourth CNN layer results are not included, as its architecture is similar to the previous layers.

This behavior becomes clearer when considering the effective stride of a multi-channel convolution after unrolling into an equivalent 1-D form. For instance, unrolling the second $a_{shape}$ layer, which has 128 channels, kernel length 8, and stride 1, produces a convolution with an effective kernel length of 1024 and an effective stride of 128. Applying the decomposition algorithm then reduces the computation to an 8-point unit stride convolution. Under these conditions, a 256-point FFT becomes highly inefficient because almost all of the FFT inputs are zero-padded.

This analysis shows that FFT efficiency depends jointly on the kernel length and the effective stride. Large FFT sizes work well for long kernel convolutions with small stride, but become inefficient when the effective stride is large or the kernel is short relative to the FFT size, since padding overhead becomes the dominant cost. Based on these observations, the 256-point FFT is selected as it provides the best trade-off between hardware complexity and performance for the bottleneck layer (the first $a_{shape}$ layer).

\begin{figure}[!ht]
    \centering
    \includegraphics[width=\columnwidth]{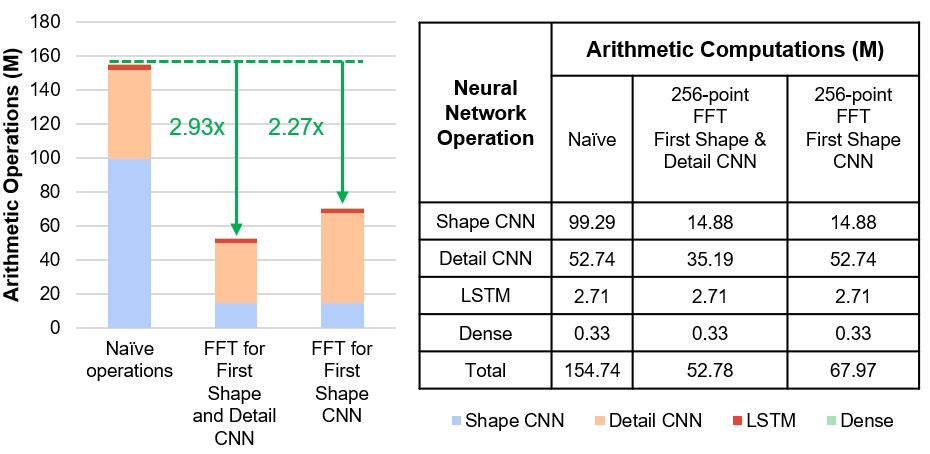}\vspace{-3mm}
    \caption{Comparison of total real arithmetic operations using naive computation and 256-point FFT fast convolution. Using the 256-point FFT achieves up to a 2.93$\times$ reduction in total operations, with
    short-kernel convolution layers becoming the new bottleneck.}
    \label{fig:FFT_comp_op}
\end{figure}

Arithmetic operations for CNN inference account for the majority of computations in the model. Notably, the first CNN layers of $a_{shape}$ and $a_{detail}$ contribute 71.5\% of the overall computation. These layers maintain performance when computed using a 256-point FFT, making them suitable candidates for optimization.
The use of FFT significantly reduces the number of real arithmetic operations required for CNN computation. As shown in Fig.~\ref{fig:FFT_comp_op}, the most computationally intensive CNN layer involves approximately 87M arithmetic operations when computed naively. By applying FFT, $a_{shape}$ can be reduced to around 4M operations.
With FFT, the two most computation-heavy layers now account for only 23\% of the model’s total operations. Overall, the adoption of a 256-point FFT reduces the total number of computations in the model by approximately 2.27$\times$.

One trade-off introduced by FFT is increased model size due to storing the
stride-based decomposed weights. The 16-bit quantized model without FFT occupies 3~MB. When FFT is applied only to the first layer of $a_{shape}$ CNN, the model size increases to 3.5~MB, representing a 16.7\% increase in exchange for a 2.27$\times$ reduction in computations, a trade-off that is well justified given the savings in arithmetic operations. Extending FFT to the $a_{detail}$ layer further reduces the computation by 22\%, at the cost of increasing the DL model size to 4.5~MB (approximately 29\%). In this case, the trade-off is not justified, as the additional reduction in computations does not outweigh the growth in model size. Therefore, our final design adopts a 256-point FFT applied only to the first layer of $a_{shape}$, which provides the best balance between computational savings and model size.
This finalized 3.5~MB model size is comparable to, and in many cases smaller than, those reported for temporally contextual sleep staging models in the literature~\cite{supratak2017deepsleepnet,mousavi2019sleepeegnet,supratak2020tinysleepnet}.

\begin{figure}[!ht]
    \centering
    \includegraphics[width=\columnwidth]{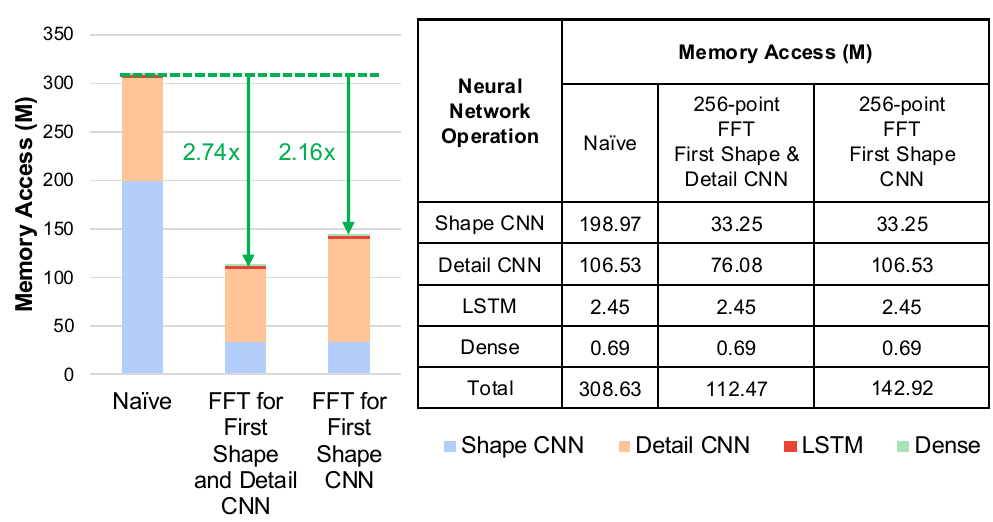}\vspace{-3mm}
    \caption{Comparison of total memory accesses for naive computation and 256-point FFT-based fast convolution applied to different layers.
    }
    \label{fig:FFT_mem_op}
\end{figure}

Reducing the number of computations not only lowers the power consumption by decreasing the workload on the MAC units, but also reduces the number of memory accesses. The table in Fig.~\ref{fig:FFT_mem_op} summarizes the number of operations required for accessing weights and data on chip. When applying the 256-point FFT only to the first layer of $a_{shape}$, the total memory access is reduced by approximately 165M.

\subsection{Quantization Testing Results}

SQNR is commonly used to assess the impact of the quantization error on model performance~\cite{lin2016fixed}. For CNNs with Gaussian-distributed weights and biases, the SQNR efficiency is typically estimated at around 5~dB per bit~\cite{you2010audio}. Previous studies have shown that convolution operations after quantization yield SQNR values between 2 to 4~dB per bit~\cite{lin2016fixed}. SQNR was evaluated over 100 randomly sampled training examples. With 16-bit PoT quantization, the average SQNR for the first CNN layer is 63.0~dB for $a_{shape}$ and 63.7~dB for $a_{detail}$, consistent with expectation. Under 8-bit PoT quantization, these values drop to 18.2~dB and 18.1~dB, respectively, remaining within the anticipated range. For BiLSTM layers, the average SQNR under 16-bit quantization is approximately 59.6~dB for a single BiLSTM layer, dropping to 22.0~dB under 8-bit quantization.
While both configurations fall within acceptable SQNR bounds, 8-bit PoT quantization causes an accuracy drop of approximately 12.2\%, whereas 16-bit PoT quantization preserves full floating-point accuracy. Accordingly,
16-bit PoT quantization is adopted for the final implementation.

For the 256-point FFT-based convolution path, quantization noise originates from the FFT, frequency-domain multiplication, and IFFT stages. Rather than applying a uniform fixed scale across all stages, scaling parameters are selected based on the statistical distribution of the training data, preventing noise accumulation while retaining the necessary dynamic range. Under this optimized configuration, 16-bit fixed-point PoT FFT fast convolution achieves an average SQNR of 40~dB over 100 samples. 8-bit FFT quantization was not pursued, as maintaining 8-bit precision consistently across the FFT, multiplication, and IFFT stages introduced significant noise.

{As introduced in Section~II-\ref{sec:model}, the CNN-BiLSTM model was evaluated on the MASS-SS3 dataset, which comprises annotated sleep recordings from 62 subjects. A leave-two-subjects-out cross-validation strategy was adopted, resulting in 31 evaluation folds. In each fold, the model was trained and validated using data from 60 subjects and tested exclusively on unseen data from the remaining two subjects, ensuring that no data from the test subjects were included during training.} The training process employed the Adam optimizer with a batch size of 256, a learning rate of $1e-4$, $\beta_1$ of 0.9, $\beta_2$ of 0.999, and an L2 weight decay of $1e-3$ over 100 epochs.

\begin{figure}[!ht]
    \centering
    \includegraphics[width=.49\textwidth]{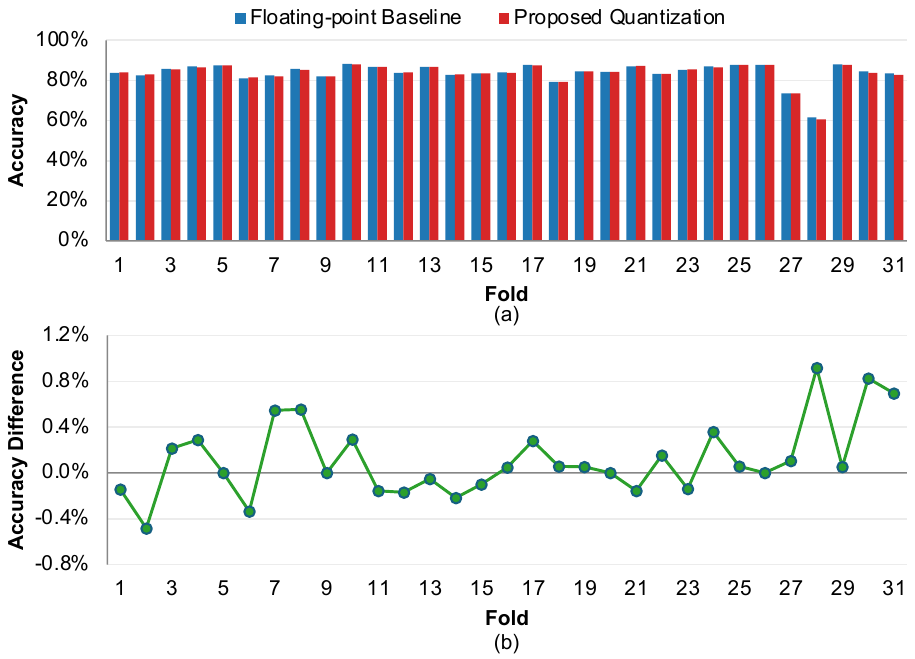} \vspace{-5mm}
    \caption{(a) Per-fold model accuracy for the floating-point baseline and after the proposed quantization. (b) Accuracy deviation of the quantized model from the floating-point baseline.}
    \label{fig:acc_difference}
\end{figure}

{
Fig.~\ref{fig:acc_difference}(a) summarizes the per-fold average accuracy of the floating-point CNN-BiLSTM model for five-class sleep-stage classification across the 31 cross-validation folds, with an overall average accuracy of 83.8\%.
The proposed 16-bit PoT quantization is applied to the trained floating-point models, and the resulting quantized accuracies are plotted alongside the floating-point baselines in Fig.~\ref{fig:acc_difference}(a). The accuracy difference between the floating-point and quantized models for each fold is highlighted in Fig.~\ref{fig:acc_difference}(b), with an average accuracy degradation of only 0.11\% across all 62 subjects. 
These results demonstrate that the proposed PoT quantization scheme preserves the subject-independent generalization capability of the floating-point model, achieving consistently negligible performance degradation across all subjects.}

\subsection{FPGA Measurement Results}

The accelerator was implemented using Xilinx Vivado 2020.2 for synthesis and implementation, and Xilinx Vitis 2020.2 for experiments. The DL model was deployed on Zynq®-7000 XC7Z020-CLG484-1 from AMD Xilinx, which integrates a Dual ARM Cortex-A9 core processor used as the MCU in this work.

\begin{figure}[!ht]
    \centering
    \includegraphics[width=0.4\textwidth]{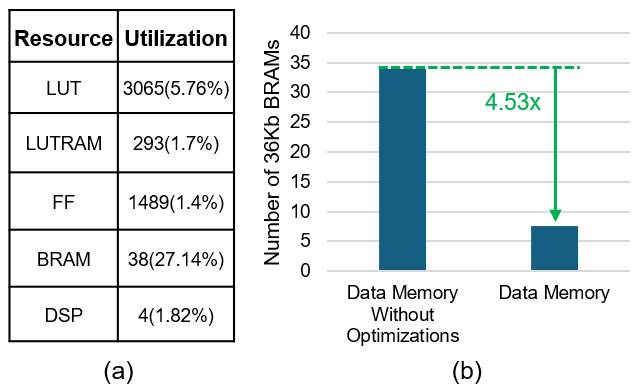}
    \caption{(a) FPGA resource utilization of complete 16-bit accelerator system including AXI interconnects on Zynq-7000 FPGA. BRAM usage is the highest of all with 38\%. (b) Comparison of data memory BRAM resource usage with and without depth-first execution.}
    \label{fig:resource_chart}
\end{figure}

Fig.~\ref{fig:resource_chart}(a) shows the programmable logic resource utilization of the model after synthesis, where BRAM (36~kbit each) resource usage is the highest among all resources. This is primarily due to the need for storing intermediate results during inference. To address this, our depth-first execution architecture significantly reduced the BRAM usage of the data memory.  As shown in Fig.~\ref{fig:resource_chart}(b), this approach, combined with the block convolution strategy, led to a 4.53$\times$ reduction in BRAM resource usage.

Building on these memory optimizations, we further improved performance by modifying the design to include a dedicated weight memory clock. During our hardware experiments, we observed that the critical path was primarily constrained by the time to transfer weights to the PL. By setting the weight clock at 2.5$\times$ that of the accelerator clock frequency and scheduling the weight transfer with convolution operations, we ensured that computations dominated the critical path.

\begin{figure}[!ht]
    \centering
    \includegraphics[width=\columnwidth]{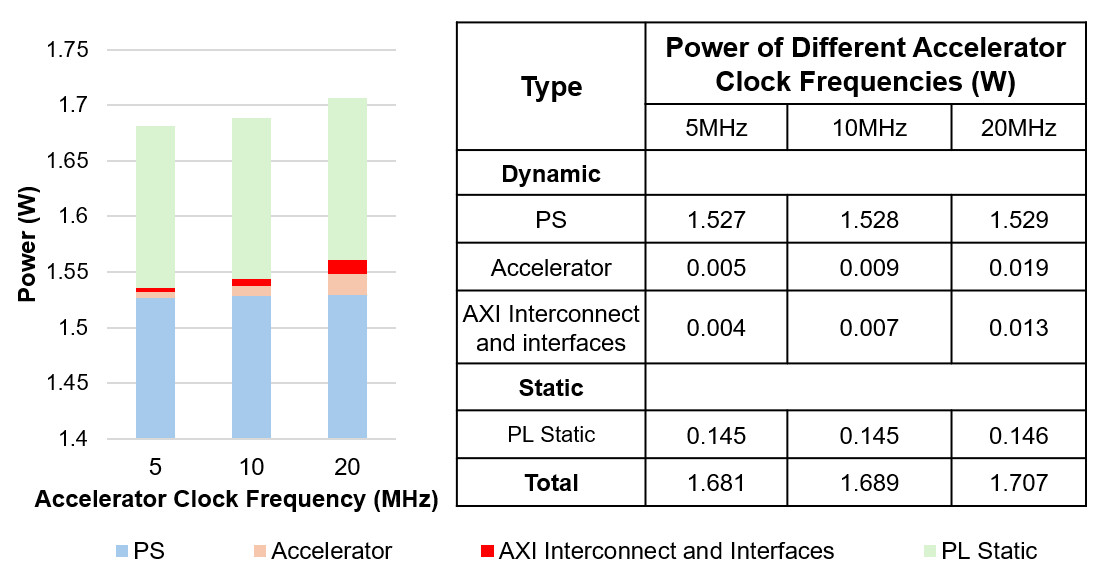}
    \caption{Power consumption of 16-bit inference on Zynq-7000 FPGA with different accelerator clock frequency from Vivado Power Analyzer.}
    \label{fig:PC_chart}
\end{figure}

To select the most energy-efficient operating point, the Xilinx Power Analyzer was used to simulate dynamic and static power across a range of clock frequencies. Fig.~\ref{fig:PC_chart} shows the power breakdown as a function of accelerator clock frequency. Since sleep stage transitions occur slowly, inference latency per sample is not a primary constraint in this application. The design that maximizes power efficiency over inference time was selected, corresponding to an accelerator clock frequency of 10~MHz and a weight memory clock frequency of 25~MHz.

Power consumption was directly measured on the Zynq-7000 FPGA.
A digital multimeter (EDU34450A, Keysight) was placed in series with the DC power supply of the FPGA development board.
The dynamic power of the accelerator was characterized by measuring the difference between operating mode with DL model inference and data movement and an idle configuration without inference and data movement. This difference was measured as 11~mW, accounting for the standalone accelerator power (matching 9~mW from simulation) as well as the MCU’s data‑handling and configuration overhead.

To evaluate hardware performance, an acceleration factor (AF) is defined as the ratio of the measured execution time to the theoretical minimum latency of conventional implementations. The theoretical minimum latency represents the latency required to perform all computations using a conventional convolution approach under ideal conditions, assuming no memory access overhead. The AF provides a measure of the effectiveness of our design in utilizing available hardware resources. An AF greater than~1 indicates an improvement over conventional limitations.

Table~\ref{model-latency} presents the experimental computation latency and the AF (the higher the better) for each layer. The FPGA was set up with a 10~MHz system clock and a 25~MHz weight memory clock.
\begin{table}[!h]
\centering
\begin{threeparttable}
\begin{center}\vspace{-0mm}
\caption{Measured Inference Latency and AF}\label{model-latency}\vspace{-0mm}
\begin{tabular}{|l|c|c|}
\hline
Modules & Latency (s)&\makecell{AF}\\\hline
First Shape Conv Layer w/ FFT &         0.617 & \textbf{7.090}\\\hline
Other Shape Conv Layers w/o FFT & 0.535 & 0.992\\\hline
Detail Conv Layers w/o FFT & 0.823\textsuperscript{*}  & 0.991\\ \hline
BiLSTM Layers&         0.360 &  0.328   \\\hline
Fully Connected Layers&         0.127&  0.135   \\ \hline
Misc. Ops (Read, Write, and Concat.) & 0.054&   N/A\\\hline
Overall Model Inference Latency&         4.162 &    1.859\\ \hline
\end{tabular}
\begin{tablenotes}
\item[\textsuperscript{*}] Three banks in the model (Fig.~\ref{fig:DL_model}).
\end{tablenotes}
\end{center}
\end{threeparttable}
\end{table}
The $a_{detail}$ and $a_{shape}$ convolution layers achieved an AF close to~1 under conventional convolution, indicating that the developed FPGA engine is highly optimized for standard convolution operations. In contrast, the FFT-accelerated layer achieved an AF of 7.09, significantly exceeding the baseline due to the use of FFT-based fast convolution. The experimental speedup is lower compared to the theoretical algorithmic reduction of 20.5$\times$. This discrepancy is primarily due to weight transfer operations occupying the critical path in the pipelined FFT convolution flow, as illustrated in Fig.~\ref{fig:FFT_schedule}. Increasing the weight memory clock frequency would mitigate this bottleneck; however, we deliberately retained the current configuration since $a_{shape}$ computations are no longer the dominant contributor to overall latency.

Similarly, BiLSTM computations are also constrained by memory transfer rates. However, they represent only a minor portion of the critical path, hence, we did not modify the memory clock frequency. Overall, our experimental inference achieves an end-to-end AF of 1.86. This result highlights the effectiveness of our FFT-based convolution engine, which leverages algorithmic acceleration to reduce arithmetic complexity and deliver high throughput with minimal hardware resources.

\section{Discussion}

{
Table~\ref{tab:soa-hardware-staging} compares this work with prior sleep staging hardware accelerators. This work achieves a competitive overall model accuracy using a CNN-BiLSTM architecture with PoT quantization, matching or exceeding prior decision-tree, filtering-based, CNN, and vision-transformer approaches (78.9\%–83.8\%) while introducing several distinct hardware-efficient acceleration techniques.}
\begin{table}[!ht]
  \centering
  \footnotesize
  \renewcommand{\arraystretch}{1.2}
  \setlength{\arrayrulewidth}{1pt}
  \setlength{\tabcolsep}{2pt}
  \caption{Comparison with Prior Sleep Staging Accelerators}
  \begin{tabular}{@{}
        >{\arraybackslash}p{1.2cm}
        >{\centering\arraybackslash}p{1.35cm}
        >{\centering\arraybackslash}p{1.35cm}
        >{\centering\arraybackslash}p{1.2cm}
        >{\centering\arraybackslash}p{1.2cm}
        >{\centering\arraybackslash}p{1.2cm} @{}}
    \toprule
                     & Imtiaz \textit{et al.}\cite{imtiaz2017ultralow} & Kassiri \textit{et al.}\cite{kassiri2016electronic} & Liao \textit{et al.} \cite{liao2021lightsleepnet} & Robitaille \textit{et al.} \cite{robitaille2026sleepvit} & \textbf{This Work} \\\midrule
    Journal          & JSSC           & TBioCAS                    & TCAS-II    & NEJ        & OJCAS      \\
    Model            & Decision tree  & Filtering \& threshold & CNN        & Vision Transformer        & CNN + BiLSTM   \\
    Dataset          & MASS           & 9 mice                     & Sleep-EDF  & MASS       & MASS       \\
    Input            & EEG            & 2 EEG + EMG                & EEG        & EEG        & EEG        \\
    Quant. Scheme & 24-bit\newline   & 8-bit \&\newline threshold         & N/A        & 16-/8-bit        & PoT        \\
    Stage \#        & 5          & 3          & 5          & 4          & 5 \\
    Accuracy         & 78.9\%         & 79.7\%                     & 83.8\%     & 82.9\%     & 83.7\%     \\
    \bottomrule
  \end{tabular}
  \label{tab:soa-hardware-staging}
\end{table}

Table~\ref{table:soa-comparison} presents a comparison with recent FPGA-based DL accelerator designs that are not limited to sleep staging applications. Prior works have traditionally emphasized standalone metrics such as GOPS/W or GOPS/DSP. However, absolute GOPS/W is strongly influenced by platform scaling on high-end devices, where higher DSP parallelism and clock frequency amortize fixed system overheads over a much larger operation count, making cross-platform comparisons misleading. To provide a more balanced assessment, we introduce DSP power efficiency (GOPS/W/DSP), which jointly captures throughput, power consumption, and is better suited for evaluating designs under tight resource constraints.

\begin{table*}[!ht]
\centering
\caption{Comparison with State-of-the-Art FPGA-based DL Accelerator Designs}
\begin{threeparttable}
\begin{tabular}{|c|c|c|c|c|c|c|c|c|}
\hline
  & \makecell{\cite{liang2019evaluating}\\TCAD\\2019} & \makecell{\cite{liang2021fcnnlib}\\TCAD\\2021} &\makecell{\cite{chang2021mixed}\\TCAS-I\\2021} &  \makecell{\cite{meng2025efficient}\\TCAS-I\\2025} & \makecell{\cite{feng2022efficient}\\{TBioCAS}\\{2022}} &\makecell{\cite{peng2023design}\\TBioCAS\\2023} & \makecell{\cite{mohamed2023unified}\\TC\\2023} & \makecell{\textbf{This} \\\textbf{Work}}\\\hline

Type& \multicolumn{4}{|c|}{ \makecell{General Purpose CNN \\Fast-Convolution Based Accelerators}}& \makecell{{Custom CNN} \\{Accelerator}} &\makecell{Custom CNN \\Accelerator} & \makecell{LSTM \\Accelerator} & Hybrid\\
\hline
Model& VGG-16&  VGG-16&VGG-16&  VGG-16& \makecell{{Custom}\\{CNN}} &\makecell{Custom\\CNN} &\makecell{Custom\\LSTM} & \makecell{Custom\\CNN+BiLSTM}\\
\hline
\makecell{Convolution\\Algorithm} & radix-2 FFT&  Winograd&Sparsity&  \makecell{Winograd+\\Sparsity}& {Conventional} &Conventional&-& FFT\\\hline
 \makecell{Bit Width (Act/Wgt)}& 8/8& 16/16& 8/8&  8/8& {(8\&16)/8} &8/8&16/16&16/16\\\hline
 Device& ZC706& ZC706& ZCU102&  VCU118& {XC7K325T} &XC7A35T& XC7Z020 &XC7Z020 \\\hline
 \makecell{Clk Freq. (MHz)} & 166& 200& 300&  300& {100} &150&100&10\\ \hline
 LUT& 350k& -& 72k& 601k& {2724} &6480\textsuperscript{~(1)}& 36.3k&3065\\\hline
 BRAM& 1090& -& 851& 792& {19.5} &20\textsuperscript{~(1)}& 82.5&38\\\hline
 DSP \#& 900& 900& 654& 1024& {21} &21\textsuperscript{~(1)}& -&4\\\hline
 \makecell{Dynamic\\Power (mW)}& 23.3k\textsuperscript{~(2)}& 7.4k\textsuperscript{~(2)}& 23.1k\textsuperscript{~(2)}& 24.4k\textsuperscript{~(2)}& {135} &250& 438\textsuperscript{~(3)}&{11\textsuperscript{~(4)}}\\\hline
 \makecell{Power Efficiency\\(GOPS/W)} & 12.74& 86.22& 16.97& 125.6&  {-} &30.28& 4.1&{3.64}\\ \hline
 \makecell{DSP Power Efficiency\\(GOPS/W/DSP)} & 0.01& 0.10& 0.03& 0.12&  {-} &0.39& -&{0.91}\\\hline
\end{tabular}
\label{table:soa-comparison}
\begin{tablenotes}
\item[\textsuperscript{(1)}] Only accelerator elements are considered.
\textsuperscript{(2)} Methodology not specified, assumed simulated and total power. Dynamic power derived from subtracting static power simulated from Vivado Power Estimator UltraScalePlus 2023.1.2 and 7-Series 2019.1.2 using resources listed in table.
\item[\textsuperscript{(3)}] Simulated MCU Power consumption removed. {\textsuperscript{(4)} Results from measurement including control and data handling power from MCU. }
\end{tablenotes}
\end{threeparttable}
\end{table*}

\textit{Ours vs. CNN Accelerators:} Table~\ref{table:soa-comparison} includes both general-purpose CNN accelerators that employ fast convolution algorithms or sparsity~\cite{liang2019evaluating, liang2021fcnnlib, chang2021mixed, meng2025efficient} and custom CNN accelerators designed for specific biomedical models~\cite{feng2022efficient, peng2023design}. General-purpose designs target high-throughput processing of large-scale models, with DSP counts ranging from 654 to 1024 and dynamic power up to 37.6~W. Custom biomedical accelerators operate at lower resource levels but still consume 135--250~mW using conventional convolution without algorithmic acceleration. Across both categories, DSP power efficiency ranges from 0.01 to 0.39~GOPS/W/DSP. The proposed design achieves 0.91~GOPS/W/DSP using only 4 DSPs and 11~mW of dynamic power, outperforming all listed CNN accelerators. This result demonstrates that FFT-based acceleration of large-kernel non-unit stride convolutions, combined with unified engine reuse and depth-first execution, enables substantially higher per-DSP efficiency than both high-resource general-purpose and task-specific biomedical designs.

\textit{Ours vs. LSTM Accelerators:} While few works specifically target LSTM acceleration, one notable work employs a CORDIC-based approach that eliminates DSP usage entirely~\cite{mohamed2023unified}, making DSP power efficiency comparisons inapplicable.
For a fair evaluation, we analyze standalone BiLSTM execution: the BiLSTM module completes inference in 360~ms for 2.368M operations, corresponding to a power efficiency of 0.60~GOPS/W. While this is lower than the CORDIC-based design, BiLSTM operations account for less than 2\% of the total inference workload. The design prioritizes minimal area overhead and hardware reuse over peak BiLSTM throughput, which is the appropriate trade-off in a system where CNN computation dominates. This highlights the advantage of system-level co-design over dedicated single-module acceleration for low-power biomedical applications.

The Zynq-7000 FPGA used in this work serves as a proof-of-concept platform for validating the proposed architecture and algorithm-hardware co-design, rather than as a final deployment target for wearable applications.
Low-power FPGA families such as Microchip IGLOO2 and SmartFusion2 provide sufficient logic and memory resources to support the proposed accelerator.
Paired with a lightweight controller, the design is feasibly deployable on such devices, where system-level power consumption is compatible with wearable operation constraints.

\section{Conclusion}
This paper presents an FPGA-based CNN-BiLSTM accelerator for real-time,
low-power sleep stage classification using single-channel EEG. The central
challenge addressed is the inefficiency of existing accelerators on
long-kernel, non-unit stride convolutions. Applying a conventional FFT
pipeline in this regime yields poor efficiency, as only 1 out of every $S$ outputs
is required. 
{
By performing stride-decomposed sub-convolution accumulations directly in the frequency domain, the proposed method achieves a 20.5$\times$ arithmetic reduction on the bottleneck layer and an overall 2.27$\times$ reduction in arithmetic complexity and 2.74$\times$ reduction in memory accesses.
This convolution-like accumulation structure is further exploited to unify the
CNN and BiLSTM compute engines, reducing hardware redundancy. Our depth-first execution further reduces the intermediate memory requirement by 4.53$\times$. Combined with block convolution and 16-bit PoT quantization, the
full system consumes 11~mW and incurs only {0.11\% average} accuracy degradation relative to the floating-point baseline.}
The accelerator, implemented on a
Xilinx Zynq-7000 FPGA at 10~MHz, achieves a 1.86$\times$ end-to-end
speedup over conventional theoretical limitations and a DSP power efficiency of
0.91~GOPS/W/DSP with 4~DSPs, outperforming reported CNN, LSTM, and hybrid
FPGA accelerators. The stride-based decomposition and quantization framework is not specific
to sleep staging and generalizes to other long-kernel time-series
workloads, such as ECG and EMG analysis, where similar kernel-size and
stride conditions arise. The modular architecture supports future
extensions including multi-channel EEG, adaptive quantization, ASIC
migration, and integration into closed-loop neuromodulation systems.

\bibliographystyle{IEEEtran}
\bibliography{ref}

\begin{IEEEbiography}[{\includegraphics[width=1in,height=1.25in,clip,keepaspectratio]{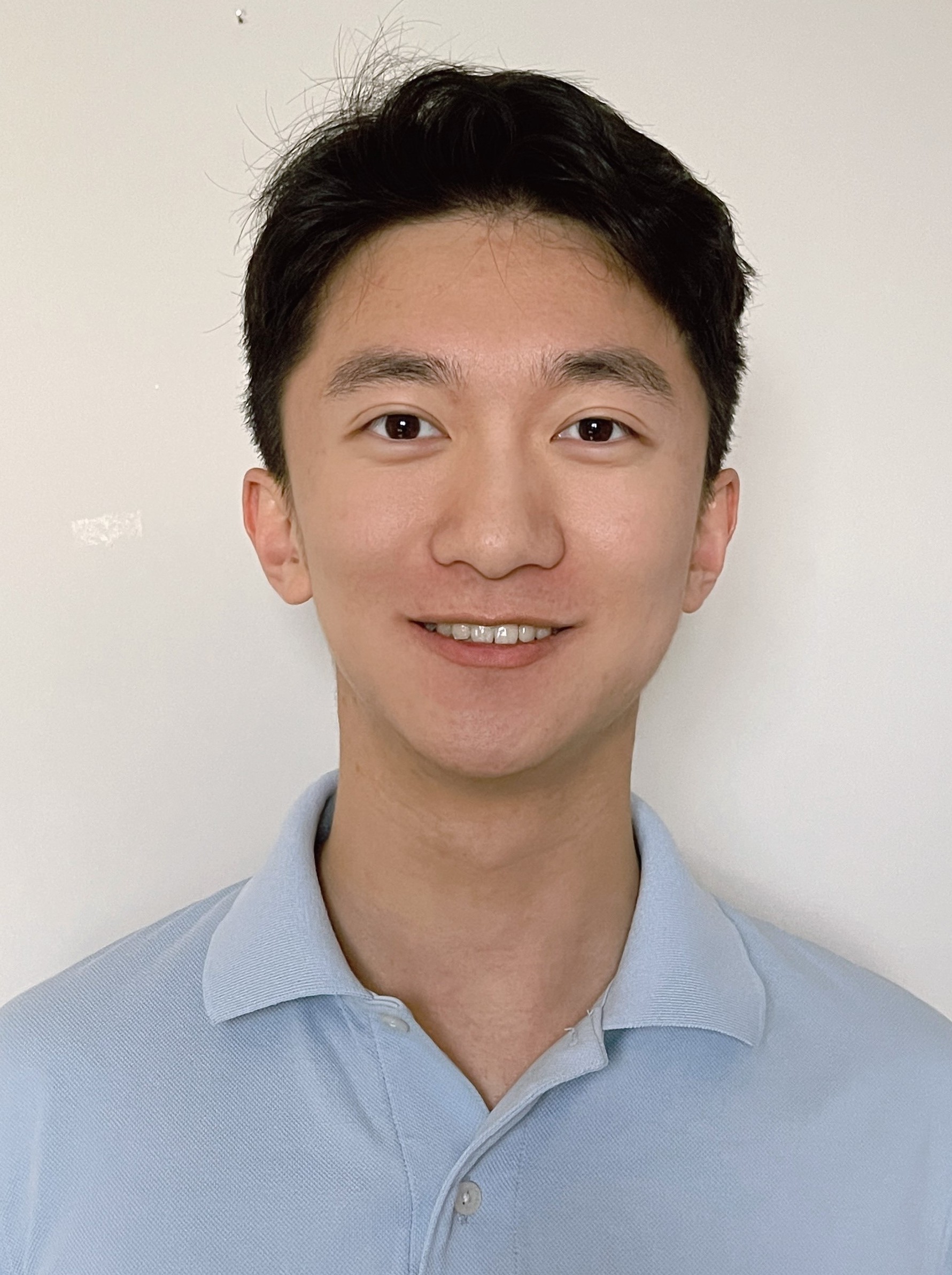}}]{Aaron Zhou}~is currently working towards the B.A.Sc. degree in Computer Engineering at University of Toronto. He joined X-Lab in 2022, where he worked on software and hardware designs for biomedical applications. His research interests include co-design of algorithms and specialized hardware for real world applications.
\end{IEEEbiography}

\begin{IEEEbiography}[{\includegraphics[width=1in,height=1.25in,clip,keepaspectratio]{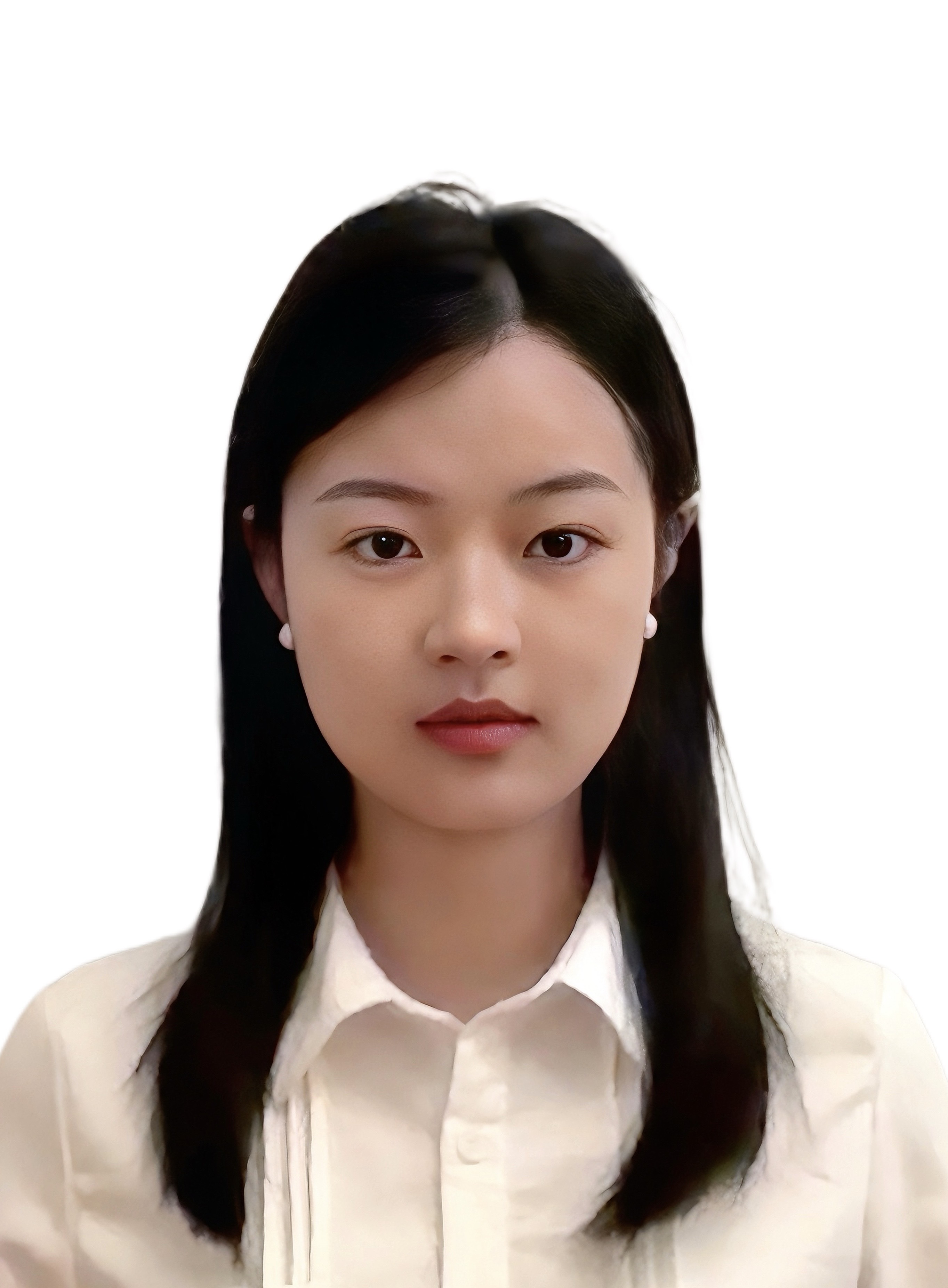}}]{Yaqian Xu}~is currently working towards the Ph.D. degree in Department of Electrical and Computer Engineering (ECE) at University of Toronto. She obtained her B.A.Sc degree in Telecommunication Engineering from Beijing University of Posts and Telecommunications and M.A.Sc degree in Biomedical Engineering from Tsinghua University, respectively.
She joined X-Lab in 2023 working on efficient machine/deep learning and its accelerations for brain machine interface applications.
\end{IEEEbiography}

\begin{IEEEbiography}[{\includegraphics[width=1in,height=1.25in,clip,keepaspectratio]{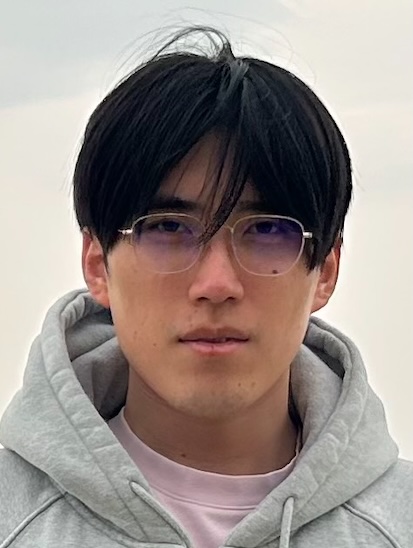}}]{Chen Chen}~received BASc degree in electrical and computer engineering from the University of Toronto, Toronto, ON, Canada, in 2024.
He is currently a Silicon Design Engineer at AMD. In 2023, he joined X-Lab at the University of Toronto, where he contributed to the development of FPGA-based hardware acceleration platforms. His interests include digital hardware design, SoC architecture, hardware acceleration, and reconfigurable systems.
\end{IEEEbiography}

\begin{IEEEbiography}[{\includegraphics[width=1in,height=1.25in,clip,keepaspectratio]{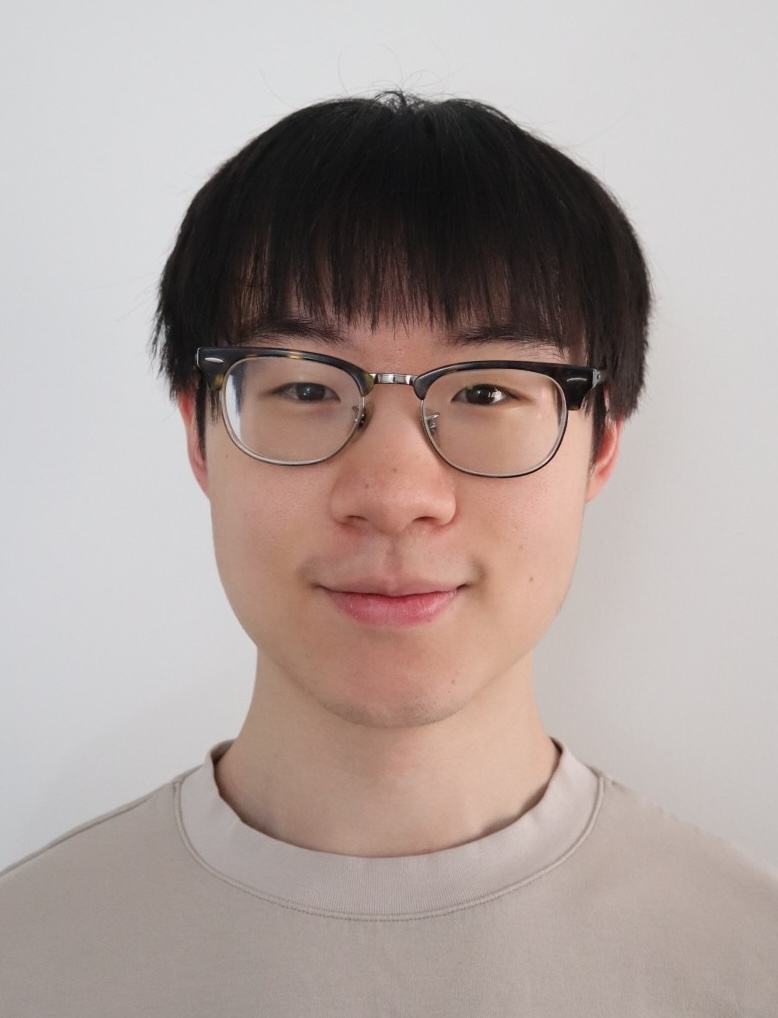}}]{James Yen}~received the B.A.Sc. Degree in Computer Engineering from the University of Toronto, Toronto, Canada. He is currently pursuing his M.A.Sc. in Computer Engineering. His research interests include heterogeneous computing and compiler design for domain-specific accelerators.
\end{IEEEbiography}

\begin{IEEEbiography}[{\includegraphics[width=1in,height=1.25in,keepaspectratio,clip]{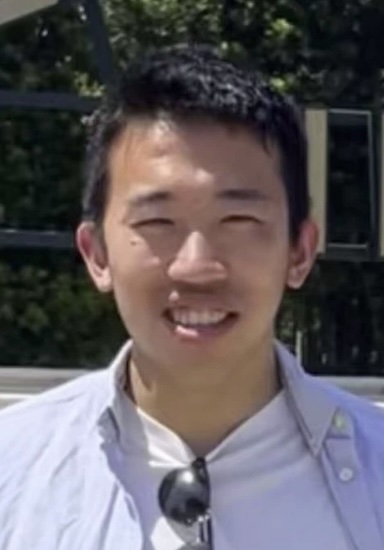}}]{Weihang Zheng}~graduated with an undergraduate degree in Computer Engineering from the University of Toronto in 2024. He now has one year of experience as an Embedded Software Developer working on ASIC drivers at Fortinet in Sunnyvale. His interests in college included various branches of ML and AI including Neural Architecture Search.
\end{IEEEbiography}

\begin{IEEEbiography}[{\includegraphics[width=1in,height=1.25in,keepaspectratio,clip]{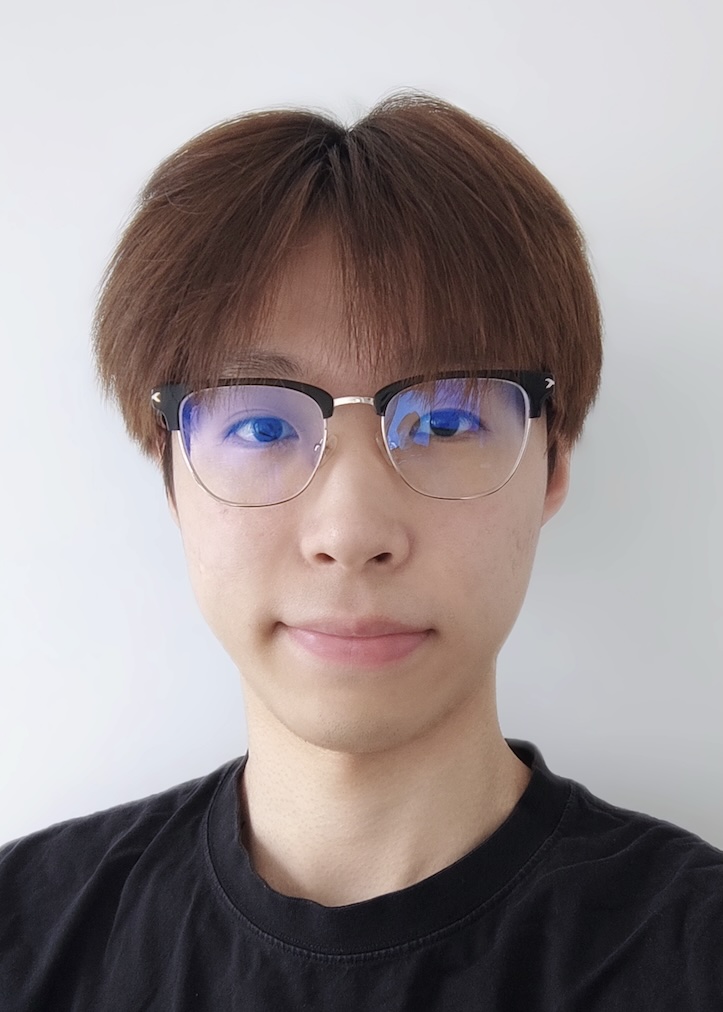}}]{Mark Zhao}~was born in Nanjing, Jiangsu, China in 2002. He received his Bachelor’s degree in Computer Engineering from University of Toronto, Toronto, Ontario, Canada, in 2025. He is currently preparing to begin his graduate studies at Georgia Institute of Technology.
\end{IEEEbiography}

\begin{IEEEbiography}[{\includegraphics[width=1in,height=1.25in,keepaspectratio,clip]{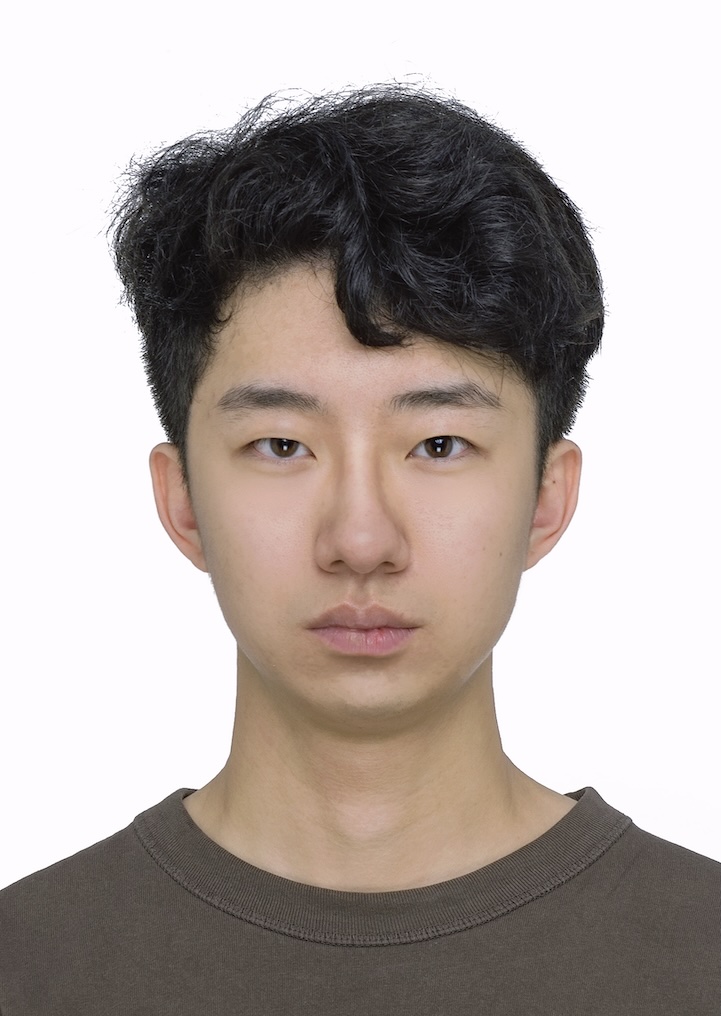}}]{Mingzhe Sun}~received his B.A.Sc. in Computer Engineering from the University of Toronto in 2023 and his M.Eng. in Electrical and Computer Engineering from Cornell University in 2024. He previously conducted research in edge machine learning at X-Lab, University of Toronto. Since 2024, he has been with TSMC as a Physical Design
Engineer, working on advanced technology nodes.
\end{IEEEbiography}

\begin{IEEEbiography}[{\includegraphics[width=1in,height=1.25in,clip,keepaspectratio]{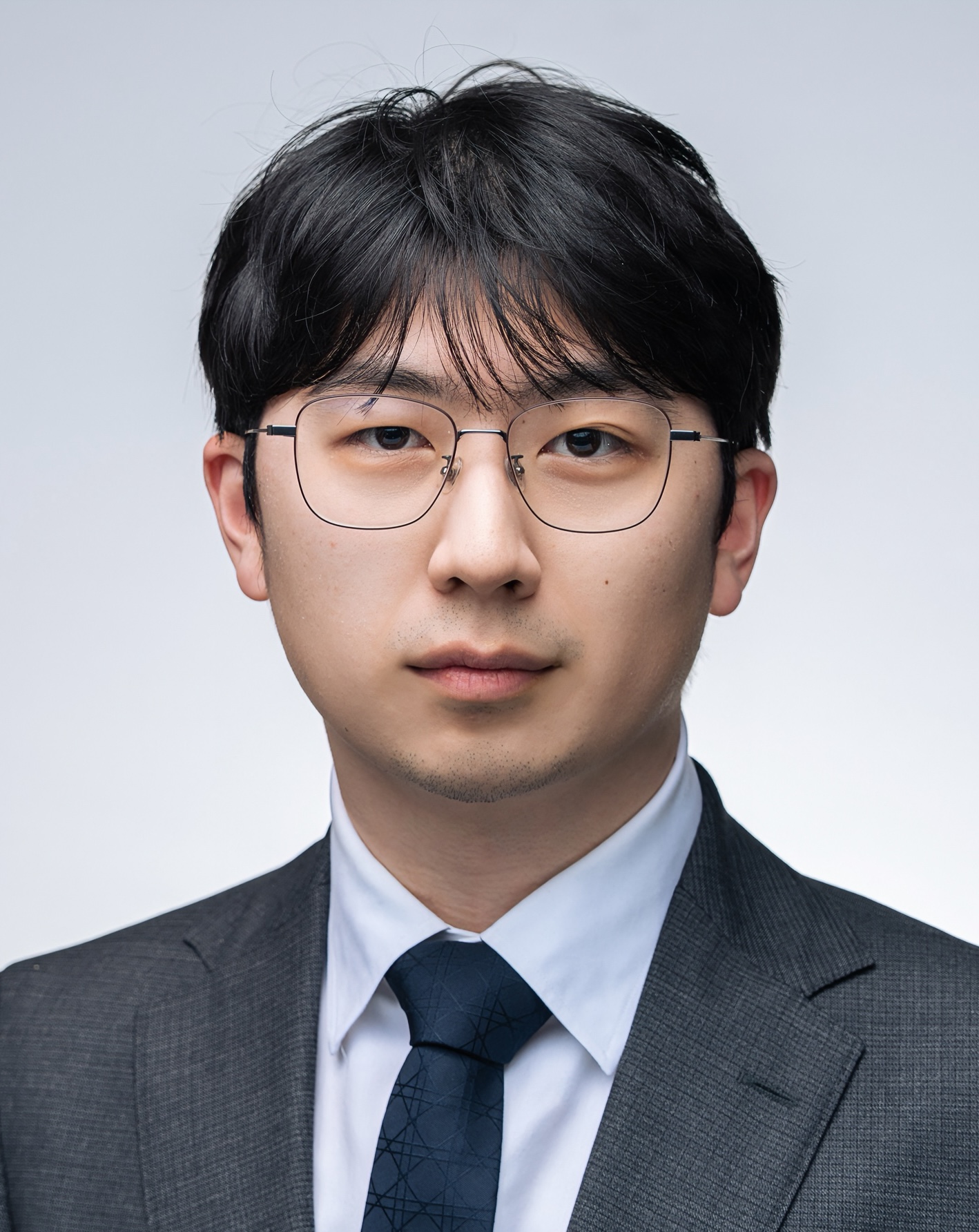}}]{Naize Yang}~(Graduate Student Member, IEEE) received the B.A.Sc. degree in Electrical Engineering from the University of Toronto, Toronto, ON, Canada, in 2023. He is currently pursuing the M.A.Sc. degree at the same institution. His research interests include High-frequency phase-locked loop (PLL) design and high-speed wireline circuit design.
\end{IEEEbiography}

\begin{IEEEbiography}[{\includegraphics[width=1in,height=1.25in,clip,keepaspectratio]{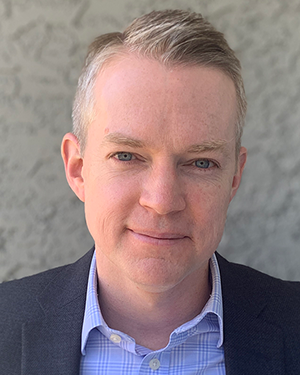}}]{Andrew G. Richardson}~(Senior Member, IEEE) received the B.S.E degree in biomedical engineering from Case Western Reserve University, Cleveland, OH, USA in 2000 and the S.M. degree in mechanical engineering and Ph.D. degree in biomedical engineering from the Massachusetts Institute of Technology, Cambridge, MA, USA in 2003 and 2007, respectively.

He was a Senior Fellow at the University of Washington, Seattle, WA, USA from 2008 to 2012 and a Senior Research Investigator at the University of Pennsylvania, Philadelphia, PA, USA from 2012 to 2015. He is currently a Research Assistant Professor of Neurosurgery at the University of Pennsylvania. His primary research interest is in brain-machine interfaces.

Dr. Richardson coauthored papers selected for Best Paper Awards at the IEEE International Symposium on Circuits and Systems in 2014 and 2017, at the IEEE Biomedical Circuits and Systems Conference in 2015, and at the IEEE Radio Frequency Integrated Circuits Symposium in 2020. His work was a finalist for the Brain-Computer Interface Award in 2018 and received an Outstanding Paper Award from the journal Microsystems \& Nanoengineering in 2021.
\end{IEEEbiography}

\begin{IEEEbiography}[{\includegraphics[width=1in,height=1.25in,clip,keepaspectratio]{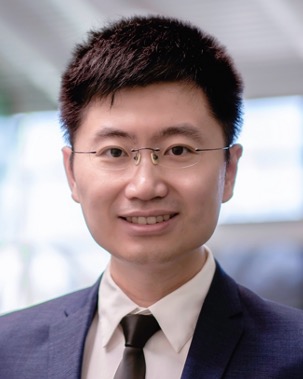}}]{Xilin Liu}~(Senior Member, IEEE) obtained his Ph.D. degree from the University of Pennsylvania, Philadelphia, PA, USA, in 2017. He is currently an Assistant Professor in the Edward S. Rogers Sr. Department of Electrical and Computer Engineering (ECE) at the University of Toronto, Toronto, ON, Canada. He is also an affiliated scientist at the University Health Network (UHN), Toronto, ON, Canada. His research interests include analog and mixed-signal IC design for biomedical circuits and systems, especially for neural interfacing and neuromodulation. Before joining the University of Toronto, he held industrial positions at Qualcomm Inc., San Diego, CA, USA.

Dr. Liu served as an Associate Editor of the \textsc{IEEE Transactions of Biomedical Circuits and Systems} in 2023-2025 and the \textsc{IEEE Transactions on Circuits and Systems II: Express Briefs} in 2023-2025. He also served on the committees of several CASS and SSCS conferences.
\end{IEEEbiography}

\vfill\pagebreak

\end{document}